# Unraveling the Origin of Ferrimagnetic Signatures in $(Fe,Mn,Ga)_2O_3$ Bixbyites: The Role of Structurally-Undetectable Spinel Impurities

Evgeniya Moshkina[a,*], Yuriy Knyazev[a], Ekaterina Smorodina[b], Oleg Bayukov[a], Maxim Molokeev[a,b,c], Evgeniy Khramov[d], Andrey Kartashev[a,e], Ruslan Batulin[f], Mikhail Cherosov[f], Dmitriy Velikanov[a], Evgeniy Eremin[a,b,g], Mikhail Rautskii[a], Dieter Kokh[h], Mikhail Platunov[i], Leonard Bezmaternykh[a]

[a]*Kirensky Institute of Physics, Federal Research Center KSC SB RAS, 660036 Krasnoyarsk, Russia*
[b]*Siberian Federal University, 660041 Krasnoyarsk, Russia*
[c]*Far Eastern State Transport University, 680021 Khabarovsk, Russia*
[d]*National Research Centre "Kurchatov Institute", Moscow 123182, Russia*
[e]*Professor V.F. Voino-Yasenetsky Krasnoyarsk State Medical University, Krasnoyarsk, 660022, Russia*
[f]*Institute of Physics, Kazan Federal University, 420008 Kazan, Russia*
[g]*Siberian State University of Science and Technologies, 660037 Krasnoyarsk, Russia*
[h]*Federal Research Center "Krasnoyarsk Science Center of the Siberian Branch of the Russian Academy of Sciences", 660036 Krasnoyarsk, Russia*
[i]*Synchrotron Radiation Facility - Siberian Circular Photon Source "SKlF" Boreskov Institute of Catalysis of Siberian Branch of the Russian Academy of Sciences (SRF "SKIF"), 630559Kol'tsovo, Russia*
*ekoles@iph.krasn.ru

**Abstract**

The cubic $Fe_{2-x}Mn_xO_3$ is an intriguing material that has recently been investigated for various applications, including lithium-ion battery anodes, catalysts, energy storage media, humidity sensors, and photocatalysts. Despite its wide range of promising applications, the magnetic properties of $Fe_{2-x}Mn_xO_3$ remain controversial, with different sources reporting conflicting information regarding the type of magnetic ordering, phase transition temperature, and magnetic moment of this compound. This work presents a study of the magnetic state of three $Fe_{2-x}Mn_xO_3$:Ga solid solutions with varying Mn:Fe:Ga ratios, along with one gallium-free $Fe_{2-x}Mn_xO_3$ reference sample. The samples were single crystals up to 3×3×3 mm³ in size, synthesized using the flux technique. We performed a detailed analysis of the actual chemical composition and crystal structure of the synthesized samples using energy-dispersive X-ray spectroscopy (EDX), powder X-ray diffraction (XRD), and X-ray absorption spectroscopy (XAS) to evaluate compositional differences. Specific heat measurements were successfully carried out on the obtained single crystals, and the observed low-temperature anomalies were analyzed in terms of magnetic phase transitions. The magnetic states of the three $Fe_{2-x}Mn_xO_3$:Ga samples and the gallium-free $Fe_{2-x}Mn_xO_3$ were investigated using magnetometry and Mössbauer spectroscopy. Mössbauer data analysis, based on a binomial distribution model, revealed differences in the local environment and site occupancy of the transition metal ions across the samples. The low-temperature magnetic anomalies were found to be more consistent with spin-glass-like freezing than with conventional long-range antiferromagnetic ordering. Although variations in magnetic behavior were observed and found to depend on composition and the cooling rate during synthesis, our results demonstrate that these factors do not account for the drastically different magnetic properties reported for similar bixbyite-type oxides. Instead, the apparent room-temperature ferrimagnetism observed in one sample is most likely extrinsic and can be attributed to a trace spinel-type impurity phase, as supported by magnetizations and ESR measurements. Thus, the origin of these discrepancies lies primarily in the chemical purity of the samples and, to a significant extent, in the synthesis technique employed.

## 1. Introduction

Due to its polymorphism, iron (III) oxide ($Fe_2O_3$) is the most interesting and promising phase among the iron oxides. The four known crystalline polymorphs of $Fe_2O_3$ – $\alpha$-, $\beta$-, $\gamma$-, and $\varepsilon$-$Fe_2O_3$ – each exhibit unique biochemical, magnetic, catalytic, and other properties, making them suitable for a wide range of technical and biomedical applications [1]. While highly crystalline $\alpha$- $Fe_2O_3$ and $\gamma$-$Fe_2O_3$ occur naturally, $\beta$-$Fe_2O_3$ and $\varepsilon$-$Fe_2O_3$ are typically synthesized in the laboratory, as are nanoparticles of all structural forms [2]. However, bulk $\beta$-$Fe_2O_3$ can be stabilized by doping with $Mn_2O_3$, resulting in the formation of the bixbyite-type solid solution $\beta$-$Fe_{2-x}Mn_xO_3$.

Like other iron oxides, bixbyite-type iron-manganese oxides $\beta$-$Fe_{2-x}Mn_xO_3$ attract considerable research interest, particularly for applications in electronics – such as negative temperature coefficient (NTC) materials [3] – as high-performance anode materials for Li-ion batteries [4], as electrode materials for energy storage [5], and even as efficient antioxidant and antibacterial agents [6]. Nevertheless, significant contradictions in the reported magnetic properties of $Fe_{2-x}Mn_xO_3$ across different studies raise concerns about the stability and/or chemical purity of the synthesized samples, despite the widespread and diverse utility of this material.

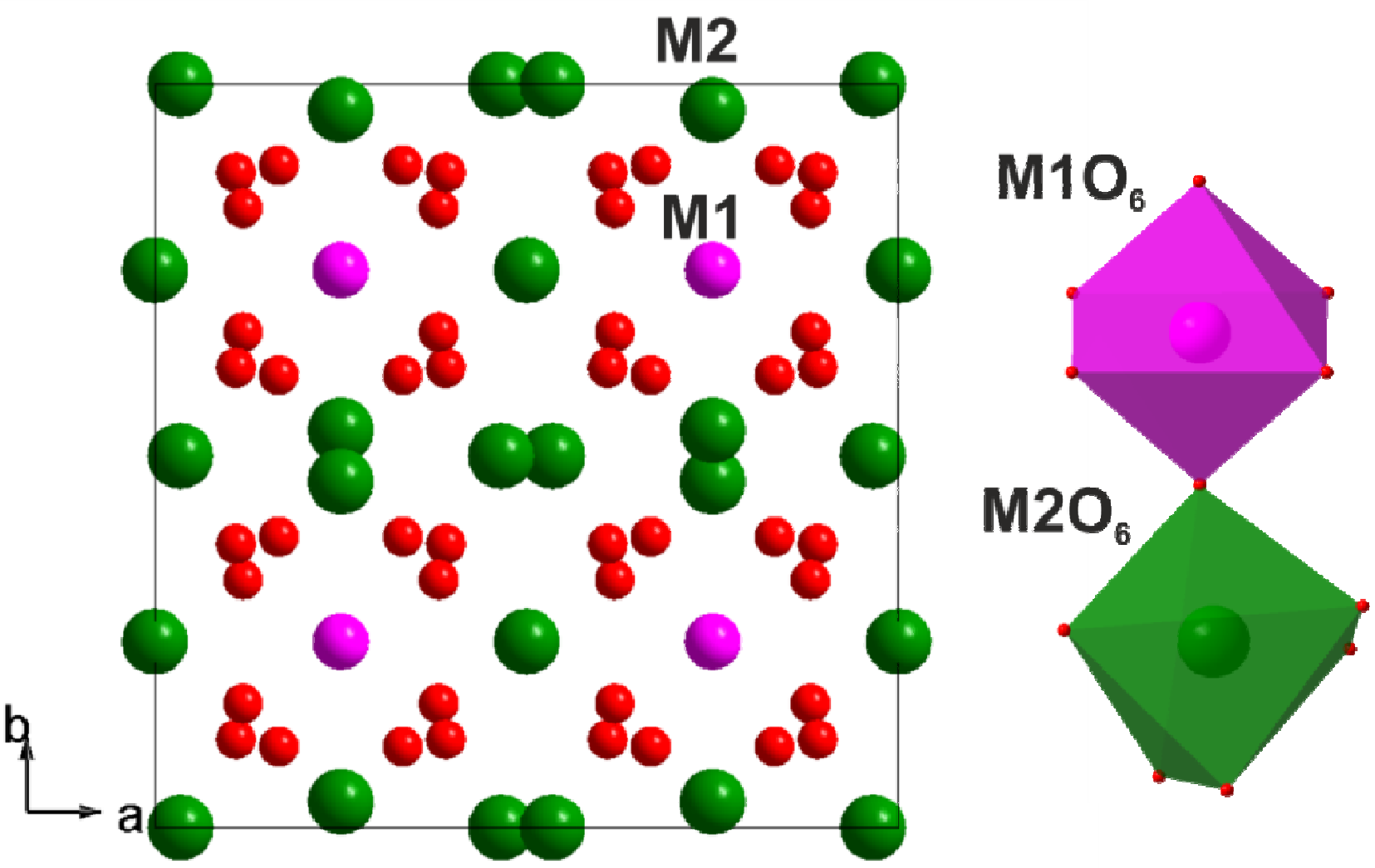

Figure 1. $Fe_{2-x}Mn_xO_3$ structure. The bixbyites have $Ia\bar{3}$ space group, the unit cell contains two formula units and two nonequivalent octahedral positions occupied by transitional metals. Purple and green spheres – M1 and M2 metal positions, respectively (M1, M2 = Mn, Fe); red spheres – oxygen.

At low temperatures, with $T_N \approx 30$–$40$ K, a magnetic phase transition is observed in many $\beta$-$Fe_{2-x}Mn_xO_3$ bixbyites, commonly described in the literature as antiferromagnetic [7–12]. However, the magnetic behavior of these materials at higher temperatures varies significantly. For instance, some studies report $FeMnO_3$ bixbyite to be paramagnetic above $T_N$ [8, 13], while others describe it as ferrimagnetic even at room temperature [9, 10, 12, 14–16].

The nature of the low-temperature phase transition itself remains controversial. According to [11], no antiferromagnetic transition occurs in $FeMnO_3$; instead, the observed anomaly is attributed to a transition into a spin-glass state, likely due to magnetic frustrations in the lattice. In contrast, [7] reports $FeMnO_3$ as antiferromagnetic below $T_N$ = 40 K. However, the temperature dependence of magnetization in that study reveals a substantial increase in magnetic moment at room temperature – more characteristic of a ferrimagnetic transition – while only weak features are observed in the low-temperature range.

A study by [12] attempted to resolve these contradictions through neutron diffraction analysis of the magnetic structure of $FeMnO_3$. Surprisingly, the results presented in that work are partially self-contradictory. The sample, a powder synthesized via mechanical alloying using a high-energy planetary ball mill, was claimed to exhibit ferrimagnetic behavior at room temperature, despite the absence of new magnetic reflections in the diffraction pattern. According to Mössbauer spectroscopy, the $^{57}Fe$ spectrum of the bixbyite remains a doublet at $T$ = 50 K – indicative of a paramagnetic state. Signs of short-range magnetic ordering first appear in the neutron diffraction data at around $T \approx 150$ K, suggesting the formation of a highly disordered magnetic phase. Upon further cooling, a distinct feature emerges at $T_N = 36$ K, which the authors associate with an antiferromagnetic phase transition. Below this temperature, [12] reports the coexistence and competition between short-range and long-range magnetic order, leading to anomalous physical and structural properties in the $FeMnO_3$ bixbyite.

This ambiguity in magnetic properties may arise from several factors, among which we highlight: (1) the cationic ordering and distribution of metal ions over inequivalent crystallographic sites (Fig. 1), and (2) the chemical purity of the synthesized samples. Both factors are strongly influenced by the sample preparation method. Given the large number of polymorphic forms of iron and manganese oxides, as well as related compounds containing these transition metals in various oxidation states, the formation of secondary phases during synthesis is highly probable. Although these impurity phases may be present in proportions too small to be detected by conventional techniques, they can still significantly affect macroscopic magnetic properties – making the second factor particularly plausible.

Recently, in the study of the multi-component flux system $Bi_2O_3$–$MoO_3$–$B_2O_3$–$Na_2O$–$Fe_2O_3$–$Ga_2O_3$, a series of Fe–Mn–Ga oxide solid solutions was obtained, three of which crystallized in the cubic bixbyite structure, $\beta$-$Fe_{2-x}Mn_xO_3$:Ga [17]. These samples exhibited different cationic compositions and Fe/Mn/Ga ratios. Thus, a series of $\beta$-$Fe_{2-x}Mn_xO_3$:Ga samples was synthesized using the same method from similar flux systems, displaying diverse magnetic behaviors that reasonably reproduce the conflicting results reported by different research groups. Studying these samples offers a unique opportunity to clarify the underlying reasons for the inconsistent magnetic properties reported for $\beta$-$Fe_{2-x}Mn_xO_3$ bixbyites. To address this, the present work employs a comprehensive set of techniques to investigate the composition, structural characteristics, cation distribution, and magnetic state of these compounds. In addition to the ternary (Fe, Mn, Ga)$_2$O$_3$ solid solutions, a gallium-free bixbyite $Fe_{2-x}Mn_xO_3$ was synthesized as a reference sample using the same flux method. The origins of the differences in magnetic behavior are systematically analyzed, with comparisons drawn based on synthesis conditions, local cationic environments, and temperature- and field-dependent magnetization and magnetic susceptibility data.

## 2. Experimental details

### *2.1. The Samples*

The experimental part of the present work presented was carried out using single-crystal samples of (Fe, Mn, Ga)$_2$O$_3$ grown by the flux method. The crystals are highly symmetrical black cubes with a maximum size of up to 3×3×3 mm³. Three samples with different Fe/Mn/Ga ratios were selected for study. Additionally, single crystals of gallium-free $\beta$-$Fe_{2-x}Mn_xO_3$ were grown using the same flux technique, from a multi-component system with the molar composition: $Bi_2O_3$ : $2.66Na_2O$ : $5.62B_2O_3$ : $0.57Gd_2O_3$ : $0.29Fe_2O_3$ : $1.45Mn_2O_3$. Based on the ratio of the crystal-forming oxides, the expected chemical formula – consistent with the flux composition – is $Fe_{0.33}Mn_{1.67}O_3$, corresponding to an Fe/Mn ratio of 1:5. The relatively low iron content compared to manganese was intentionally chosen to investigate the selectivity of $Fe^{3+}$ ion occupation at the inequivalent crystallographic sites in the $\beta$-$Fe_{2-x}Mn_xO_3$ bixbyite structure. Crystal growth was performed at a constant temperature of 920 °C over a period of one day, under conditions of spontaneous nucleation, using a platinum rod as a crystal holder. The saturation temperature of the flux was $T_{sat}$ = 930 °C. A detailed description of the growth procedure is provided in Ref. [17].

*2.2. X-ray diffraction*

Powder diffraction data for Rietveld analysis were collected at room temperature using a Haoyuan DX-2700BH powder diffractometer equipped with Cu-Kα radiation and a linear detector. The scan was performed with a step size of 0.01° in 2θ and a counting time of 20 seconds per degree. Rietveld refinement was carried out using TOPAS 4.2 [18].

*2.3. Energy dispersive X-ray spectroscopy (EDX)*

Energy dispersive X-ray spectroscopy (EDX) was employed to analyze the chemical composition of the synthesized single-crystal samples. The samples were mounted on an aluminum stub using double-sided carbon tape and then placed into the sample chamber of a Hitachi SU3500 scanning electron microscope (Japan), equipped with a Bruker XFlash 6160 energy-dispersive detector (Germany). Electron micrographs were acquired in backscattered electron (BSE) mode using an accelerating voltage of 20 kV and a cathode brightness of 60 arbitrary units. These parameters were kept constant for all recorded images to ensure consistent imaging conditions.

*2.4. X-ray absorption spectroscopy*

X-ray absorption spectra were measured in transmission geometry at the "Structural Materials Science" beamline of the Kurchatov Synchrotron Radiation Source [19]. The X-ray energy was selected using a channel-cut Si (111) monochromator. The synchrotron beam size at the sample position was 0.3 × 1 mm² (V × H). Samples were prepared as pellets by pressing finely ground powder with starch, which served as a binding agent. The sample amount was carefully chosen to ensure that the absorption coefficient (μt) did not exceed 3–3.5 across the entire energy range. The intensities of the incident and transmitted X-rays were measured using ionization chambers filled with an air-argon mixture, with current detection performed using Keithley 6487 picoammeters.

Absorption spectra were recorded point by point over an energy range from −170 eV to +800 eV relative to the K-edge absorption energies of the absorbing elements: iron (7112 eV), manganese (6539 eV), and gallium (10367 eV). Each spectrum was measured for approximately 20 minutes and repeated 2–3 times per sample; the resulting scans were averaged to improve signal-to-noise ratio. Data processing and analysis were carried out using the Athena program from the IFEFFIT software package (version 1.2.11c) [20-21].

The oxydation states of manganese and iron were determined by "fingerprint" analysis of the XANES spectra. This involved comparing the position and shape of the K-edge absorption features in the samples with those of reference compounds: $\gamma$-$Fe_2O_3$ ($Fe^{3+}$), FeO ($Fe^{2+}$), metallic

Fe, $Mn_2O_3$ ($Mn^{3+}$), MnO ($Mn^{2+}$), and metallic Ga. The spectral shape near the absorption edge and the first derivative of the absorption spectra were visually analyzed to identify inflection points, which are indicative of the oxidation state of the absorbing atom.

*2.5. Mössbauer spectroscopy*

Mössbauer spectra were recorded at room temperature using a $^{57}$Co (Rh) source and an MS-1104Em spectrometer equipped with a 512-channel analyzer. The spectral linewidth, measured with a standard $\alpha$-Fe absorber and a NaI (Tl) detector, was 0.24 mm/s. Measurements were performed on powdered samples prepared from single crystals, with a thickness of 5 - 10 mg/cm² in terms of natural iron content.

Spectral analysis was carried out in two stages. In the first stage, the probability distribution of quadrupole splitting, *P*(QS), was calculated. Features in this distribution indicated the presence of multiple nonequivalent iron sites or local environments. In the second stage, model spectra were fitted to the experimental data by varying the full set of hyperfine parameters using a least-squares fitting procedure within the linear approximation. This refinement included the parameters of individual doublets in the paramagnetic phase and sextets in the magnetically ordered phase – specifically: isomer shift, quadrupole splitting, hyperfine magnetic field, and spectral amplitude. The isomer shift values are reported relative to metallic $\alpha$-Fe.

*2.6. Specific Heat Measurements*

The experiments were carried out using an adiabatic calorimeter. Adiabatic conditions were maintained by means of a high vacuum ($10^{-6}$ – $10^{-5}$ mbar) and a temperature-controlled "tracking" screen, which simultaneously minimized direct heat exchange with the environment. A platinum thermometer was mounted on the tracking screen to monitor its temperature. The temperature difference between the sample cell and the tracking screen was monitored using a thermocouple (manganin-constantan). This setup allowed the temperature scan rate to be controlled within the range of $10^{-5}$ to $10^{-3}$ K/min without the application of external heating power. The linearity of the temperature dependence confirms the absence of significant thermal gradients. This technique enables the acquisition of a series of experimental heat capacity data over various temperature ranges. The length of each measurement series was determined experimentally, based on both the characteristics of the sample and the specific experimental conditions. The lowest heating rates ($10^{-5}$ – $10^{-3}$ K/min) are limited by the precision of temperature control, while the highest rates are constrained by the onset of thermal gradients and the operational limits of the electronic PID controller. In this study, heating rates between 0.01 and 1 K/min were employed. The measurements were performed on three polycrystalline samples with masses of 32.9 mg, 53.78 mg, and 96.00 mg, and one single-crystal sample with a mass of 39.35 mg. Apiezon N grease was used to fix the samples to the sample holder and ensure good thermal contact. The heat capacity of the empty sample cell, including the grease, was determined in a separate experiment and subtracted from the total measured heat capacity to obtain the sample contribution.

*2.7. Magnetic Measurements*

Temperature- and field-dependent magnetization and magnetic susceptibility measurements were performed on the synthesized samples over the temperature range of 4.2–420 K and in magnetic fields up to 9 T, using a PPMS-9 Vibrating sample magnetometer options (VSM and VSM Oven, Quantum Design, USA) and a custom-built Vibrating Sample

Magnetometer (VSM) and SQUID magnetometer developed at the Kirensky Institute of Physics, SB RAS [22, 23]. VSM Oven allows to extend the temperature range (2-400 K) of a standard VSM to 1000 K.

## 3. Results

The results of the study and comparison of the structural and thermodynamic properties—including magnetization and heat capacity – of three monocrystalline $\beta$-(Fe, Mn, Ga)$_2$O$_3$ samples and one gallium-free reference sample, $\beta$-(Fe, Mn)$_2$O$_3$, with different compositions obtained from various flux systems, presented in Ref. [17].

### *3.1. Structure and Chemical Composition*

The cubic bixbyite phase of the selected samples, crystallizing in the $Ia\overline{3}$ space group, was confirmed by powder X-ray diffraction. The lattice parameters are listed in Table 1. In this structure, transition metal cations occupy two inequivalent crystallographic sites: *8b* and *24d*. For the studied (Mn, Fe, Ga)$_2$O$_3$ compounds, these sites can be occupied by three types of trivalent cations: $Fe^{3+}$, $Mn^{3+}$, and $Ga^{3+}$.

The chemical composition of the samples was initially estimated using energy-dispersive X-ray spectroscopy (EDX) and subsequently refined based on the edge-jump intensities at the K-edges of iron, manganese, and gallium in the XANES spectra. The results are also summarized in Table 1. It is evident that the compositions determined by EDX differ from those obtained by XANES for all ternary solid solutions, particularly in the relative concentrations of gallium compared to iron and manganese. Since XANES provides lower uncertainty in quantification for these elements than EDX, we consider the XANES-derived compositions to be closer to the true values. Accordingly, the actual compositions of the ternary solid solutions are characterized by a higher gallium content, which increases with increasing manganese concentration, while the iron content decreases correspondingly. No XANES data are available for the gallium-free $Fe_{0.52}Mn_{1.48}O_3$ sample. Thus, the EDX-derived composition of $Fe_{0.52}Mn_{1.48}O_3$ is used.

With respect to compositional changes, the unit cell parameters of the ternary oxides vary only slightly, due to the small differences in ionic radii of the cations ($R$ ($Fe^{3+}$) = $R$($Mn^{3+}$) = 0.645 Å, $R$ ($Ga^{3+}$) = 0.62 Å [24]) and the relatively similar cationic compositions. Only the lattice parameter of the gallium-free sample shows a significant deviation, attributed to its distinct composition and the absence of $Ga^{3+}$, which has a smaller ionic radius.

Table 1. Lattice parameters of three (Mn, Fe, Ga)$_2$O$_3$ and $Fe_{0.52}Mn_{1.48}O_3$ obtained by powder X-ray diffraction. The chemical composition of the studied samples presented from EDX and XANES analysis.

| No. | Chemical formula (EDX) | Chemical formula (XAS) | $a$, Å | $V$, Å$^3$ |
|---|---|---|---|---|
| S1 | $Fe_{1.1}Mn_{0.76}Ga_{0.14}O_3$ | $Fe_{0.93}Mn_{0.68}Ga_{0.39}O_3$ | 9.40007(13) | 830.60(4) |
| S2 | $Fe_{1.07}Mn_{0.72}Ga_{0.21}O_3$ | $Fe_{0.78}Mn_{0.68}Ga_{0.54}O_3$ | 9.40024(17) | 830.65(4) |
| S3 | $Fe_{0.7}Mn_{1.08}Ga_{0.22}O_3$ | $Fe_{0.5}Mn_{0.84}Ga_{0.66}O_3$ | 9.39891(13) | 830.30(4) |
| S4 | $Fe_{0.52}Mn_{1.48}O_3$ | | 9.41744(75) | 835.21(2) |

### *3.2. Magnetic Properties*

Fig. 2 shows the temperature-dependent magnetic susceptibility of the ternary (Fe, Mn, Ga)$_2$O$_3$ samples and the gallium-free $Fe_{0.52}Mn_{1.48}O_3$ sample, measured under zero-field-cooled (ZFC) and field-cooled (FC) conditions. As can be seen from the figure, the low-temperature behavior of all samples is similar: a peak in susceptibility is observed at $T$ = 30-40 K in both measurement modes (see inset in Fig. 2), indicating a magnetic phase anomaly associated with either an antiferromagnetic phase transition or a transition into a spin-glass state, as previously reported in bixbyites [11]. However, at higher temperatures, the magnetization curves of sample S2 deviate significantly from those of the others. Below $T \approx 370$ K, a noticeable increase in susceptibility is observed, suggesting a second magnetic phase transition. This temperature range has not been previously explored in (Fe, Mn, Ga)$_2$O$_3$ compounds, indicating that this anomaly is observed in bixbyites for the first time.

At $T$ = 4.2 K, samples S1 and S2 – despite their differing iron content – exhibit nearly identical FC magnetic susceptibility values (Fig. 2). The susceptibility of sample S3 at the same temperature is slightly higher. Moreover, the overall shape of its $\chi(T)$ dependence differs from that of S1 and S2: instead of saturating at low temperatures (as seen in S1 and S2), a weak increase in magnetic susceptibility is observed below $T \approx 10$ K (Fig. 2). The temperature-dependent magnetization of the gallium-free sample S4 ($Fe_{0.52}Mn_{1.48}O_3$) behaves similarly to that of S1 and S3. However, the magnitude of the susceptibility peak at the magnetic transition in S4 is more than twice as large as that in the gallium-containing samples. This is consistent with the fact that $Ga^{3+}$ does not contribute to the magnetic moment, thereby reducing the overall magnetic response in the doped compounds.

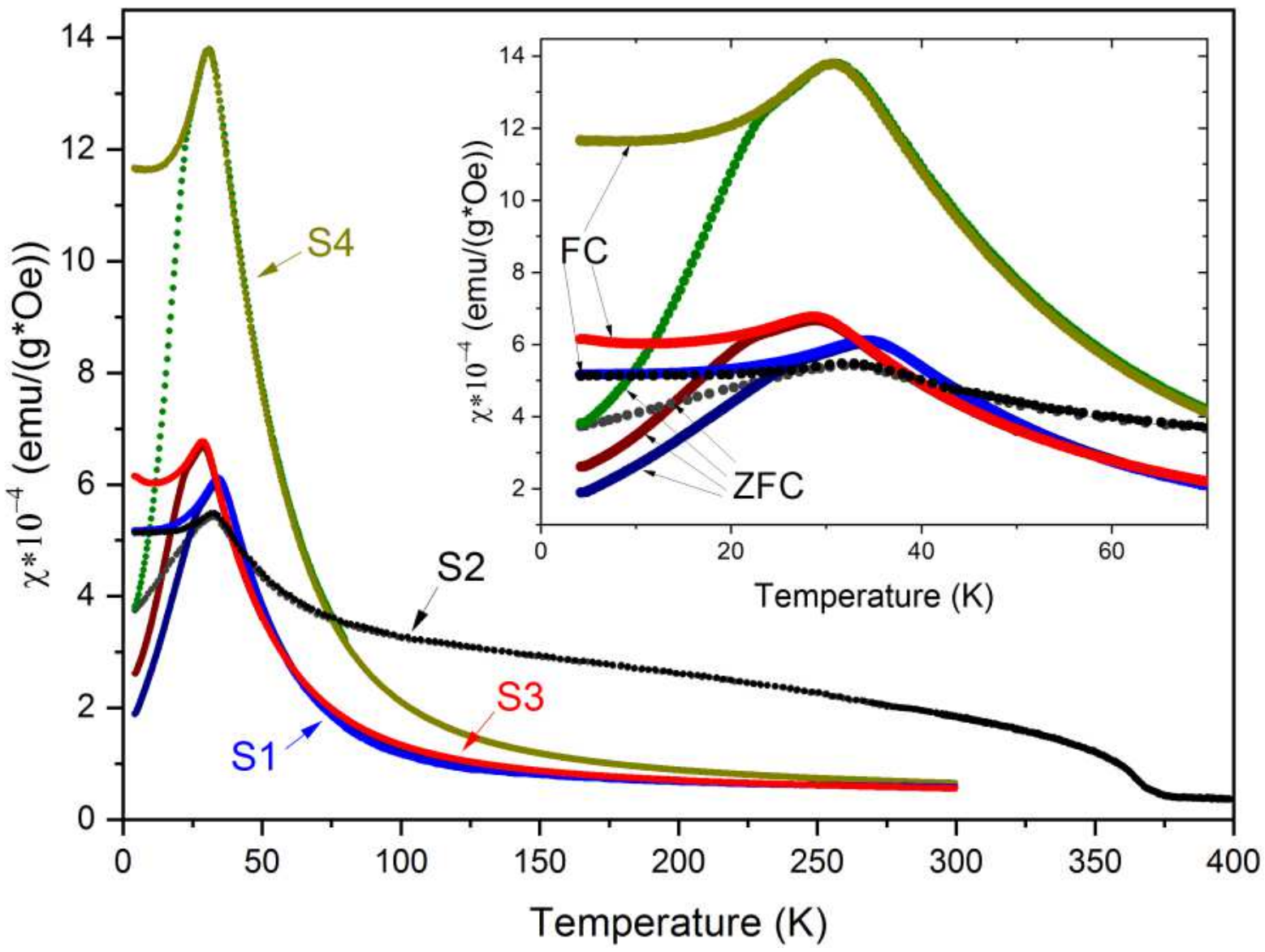

Figure 2. Temperature dependence of the magnetic susceptibility of $Fe_{1.1}Mn_{0.76}Ga_{0.14}O_3$ (S1), $Fe_{1.07}Mn_{0.72}Ga_{0.21}O_3$ (S2), $Fe_{0.7}Mn_{1.08}Ga_{0.22}O_3$ (S3) and $Fe_{0.52}Mn_{1.48}O_3$ (S4), obtained using a PPMS-9 system (S1, S3 and S4, up to 300 K), SQUID magnetometer (S2, up to 273 K), a VSM magnetometer (S2, $T$ = 290-400 K). FC – field cooling regime, ZFC – zero-field cooling regime. The magnetization data used for plotting were obtained at $H$ = 1 kOe and $H \parallel a$.

Thus, qualitatively different magnetic behavior was observed in the bixbyite (Fe, Mn, Ga)$_2$O$_3$ system. Samples S1, S3, and S4 are paramagnetic at room temperature and undergo a magnetic phase transition at approximately 30 K. In contrast, sample S2 exhibits two phase transitions: a ferrimagnetic transition at $T_C$ = 360-370 K and a low-temperature transition at $T \approx$ 30 K, similar to the other samples. The temperature-dependent magnetization curves of sample S2 are qualitatively consistent with those reported in [9–10, 12] for gallium-free bixbyites. Notably, samples S1 and S2 have comparable chemical compositions. This suggests that the emergence of the high-temperature ferrimagnetic transition either critically depends on subtle compositional differences or is influenced by a different distribution of metal cations over the inequivalent crystallographic sites (8*b* and 24*d*) within the unit cell.

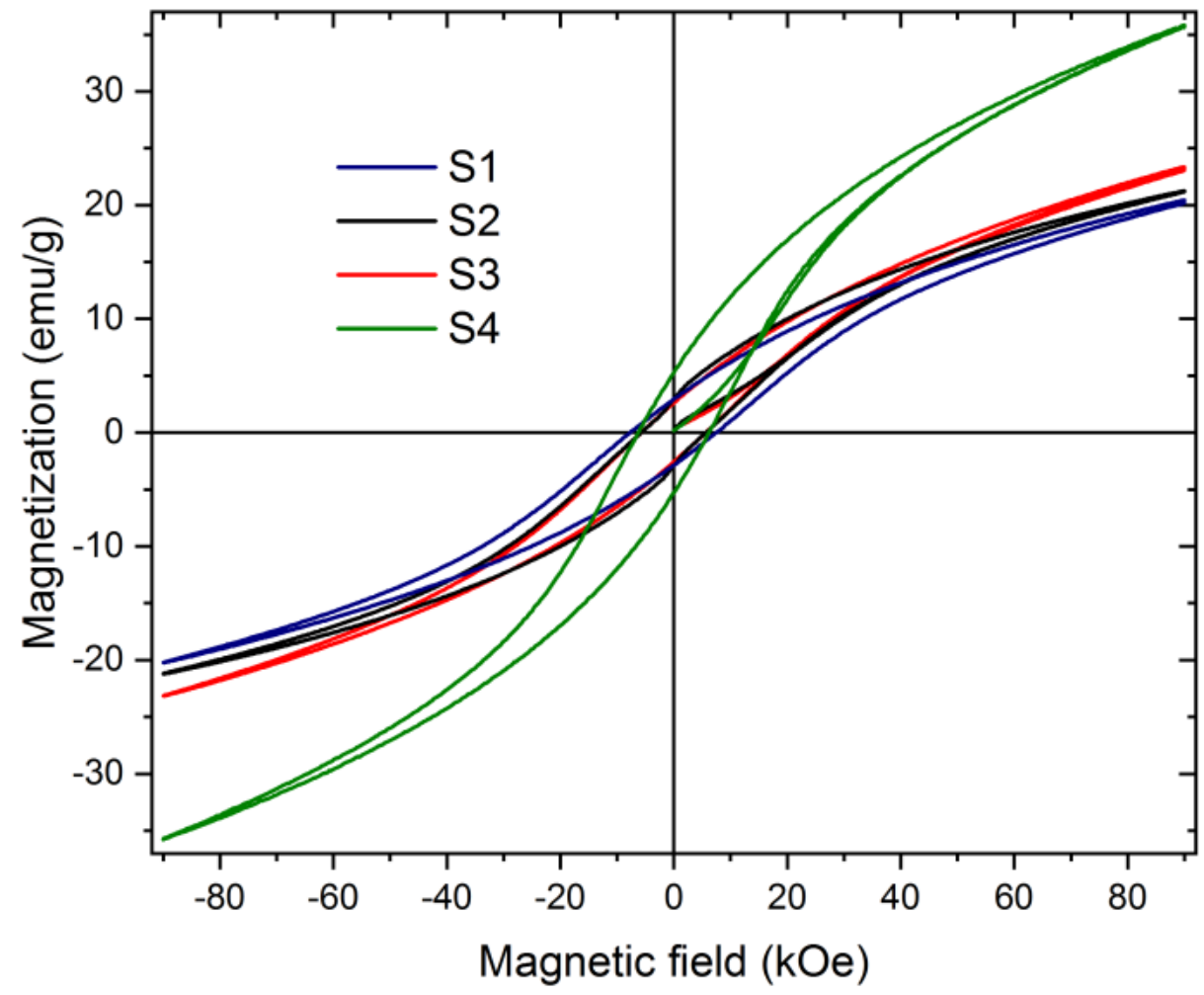

Figure 3. Field dependences of the magnetization of the studied S1, S2, S3 and S4 bixbyites. $T$ = 4.2 K, $H||a$.

The field-dependent magnetization curves of the studied samples, measured at $T$ = 4.2 K, are shown in Fig. 3. Qualitatively, the magnetic hysteresis loops – like the low-temperature magnetization behavior – are similar across all samples: they exhibit unsaturated behavior in fields up to 9 T. The coercive field is approximately $H_c \approx 0.55$ T for samples S2 and S3, and slightly higher, $H_c \approx 0.75$ T, for sample S1. The slope and saturation trend of the loops differ slightly among the samples. Among the ternary $(Fe, Mn, Ga)_2O_3$ compounds, sample S3 – which has the lowest iron content – exhibits the largest magnetic moment at 9 T. The gallium-free reference sample displays a coercive field comparable to that of S2 and S3, but its magnetization at 9 T is significantly higher – by more than 1.5 times – than that of the gallium-containing samples. This is consistent with the previously observed temperature-dependent magnetization data and can be attributed to the absence of non-magnetic $Ga^{3+}$ ions, which dilute the magnetic lattice.

### *3.2. Cation Distribution Analysis*

The studied compounds contain three types of cations – Mn, Fe, and Ga – which occupy two inequivalent crystallographic sites, 8*b* and 24*d* [25]. The X-ray atomic scattering factors of these elements are very similar, making it impossible to determine site occupancies reliably using conventional X-ray diffraction methods. However, the magnetic properties of $(Mn, Fe, Ga)_2O_3$ are highly sensitive to the distribution of these cations over the crystallographic sites. To investigate the site-specific distribution of Mn, Fe, and Ga in the studied samples, element-selective X-ray absorption spectroscopy (for the ternary oxides) and Mössbauer spectroscopy were employed. These techniques were combined with a binomial distribution analysis applied to all four samples, enabling a detailed examination of the iron cation distribution across the 8*b* and 24*d* sites.

#### *3.2.1. Mössbauer Spectra Analysis Using Binomial Distribution*

Mössbauer spectra of the studied samples are shown in Fig. 4. All spectra exhibit the form of a quadrupole doublet, indicating that iron cations are in a paramagnetic state in these crystals. Since each spectrum consists of multiple overlapping doublets, the probability distributions of quadrupole splitting, *P*(QS), were calculated using the procedure described in detail in [26-27]. The resulting *P*(QS) distributions are presented as histograms in the right panel for each sample. These distributions reveal several distinct types of local environments around the iron cations across the series, indicating a non-uniform distribution of $Fe^{3+}$ ions over the crystallographic sites. The spectral components were fitted according to the *P*(QS) distributions, and the fitting parameters are summarized in Table 2.

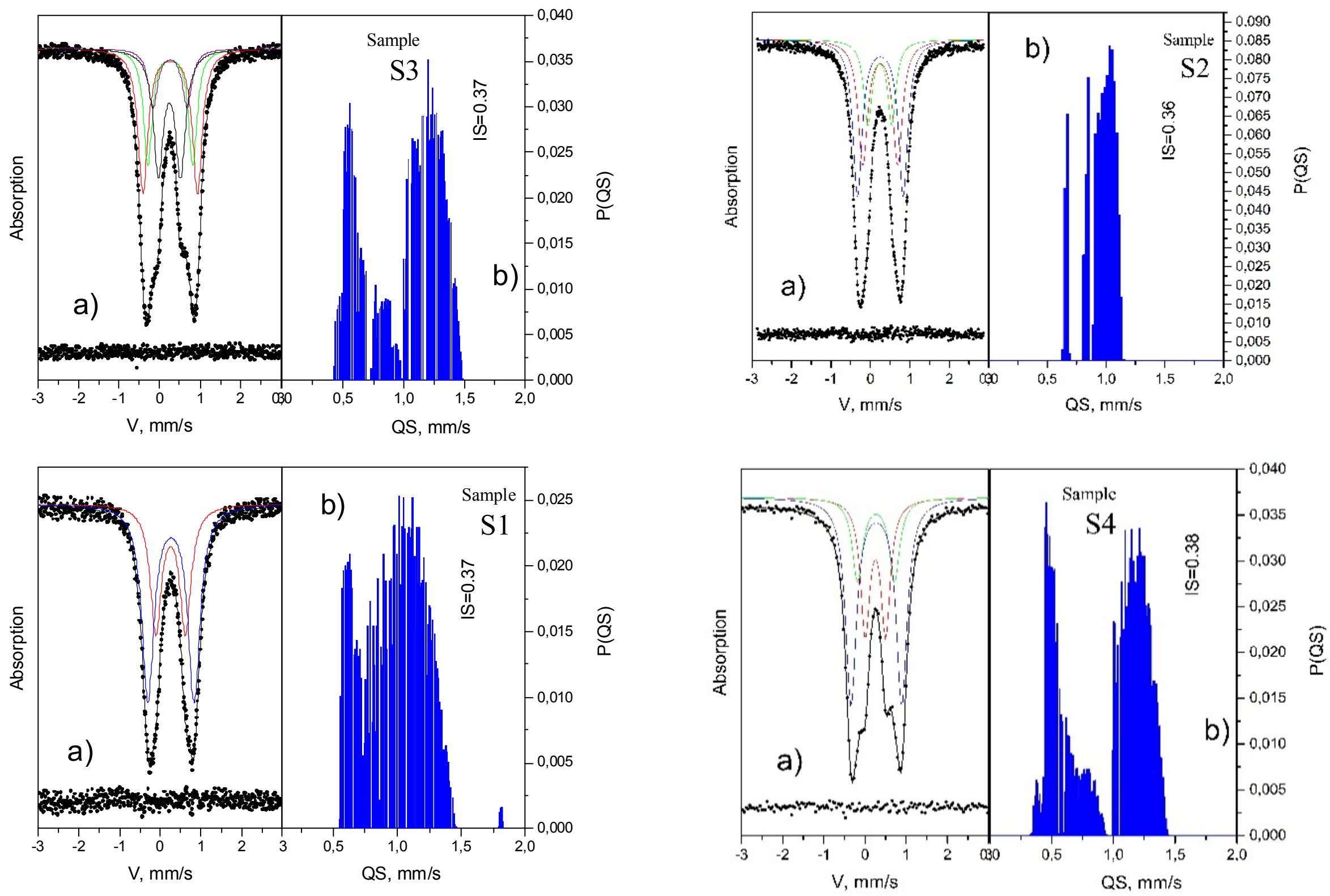

Figure 4. Mossbauer spectra of the studied samples measured in transmission geometry at *T* = 300 K. The error signal is shown below each spectrum. The sample number are indicated in the figure. The right-hand panels show the calculated quadrupole splitting probability distributions.

Table 2. Mössbauer parameters of the samples. *IS*: isomer shift relative to *α*-Fe; *QS* (2*ε*): quadrupole splitting; *W*: line width at half-height (full width at half maximum, FWHM); *Area*: relative spectral area, proportional to the relative site occupancy. D1, D2, D3, D4: individual spectral doublets.

| Sample No., Space group | № | IS, mm/s ±0.005 | QS (2ε), mm/s ±0.02 | W, mm/s ±0.02 | Area, ±0.03 | Position |
|---|---|---|---|---|---|---|
| S1 $Ia\overline{3}$ | D1 | 0.373 | 0.73 | 0.33 | 0.38 | M1 |
| | D2 | 0.385 | 1.14 | 0.36 | 0.62 | M2 |
| | | | | | | |
| S2 $Ia\overline{3}$ | D1 | 0.360 | 0.88 | 0.29 | 0.36 | M1 |
| | D2 | 0.361 | 1.16 | 0.29 | 0.44 | M2 |
| | D3 | 0.349 | 0.61 | 0.25 | 0.20 | |
| S3 $Ia\bar{3}$ | D1 | 0.361 | 0.55 | 0.30 | 0.29 | M1 |
|  | D4 | 0.375 | 0.84 | 0.28 | 0.13 |  |
|  | D2 | 0.372 | 1.35 | 0.29 | 0.34 | M2 |
|  | D3 | 0.388 | 1.11 | 0.26 | 0.24 |  |
|  |  |  |  |  |  |  |
| S4 $Ia\bar{3}$ | D1 | 0.362 | 0.50 | 0.27 | 0.29 | M1 |
|  | D2 | 0.384 | 1.23 | 0.32 | 0.51 | M2 |
|  | D3 | 0.376 | 0.90 | 0.30 | 0.20 |  |

Taking into account the Mössbauer spectroscopy data and the chemical compositions of the samples determined by X-ray absorption spectroscopy (Table 1), it is possible to estimate the probability of finding iron cations in the second coordination sphere, considering cation substitution effects.

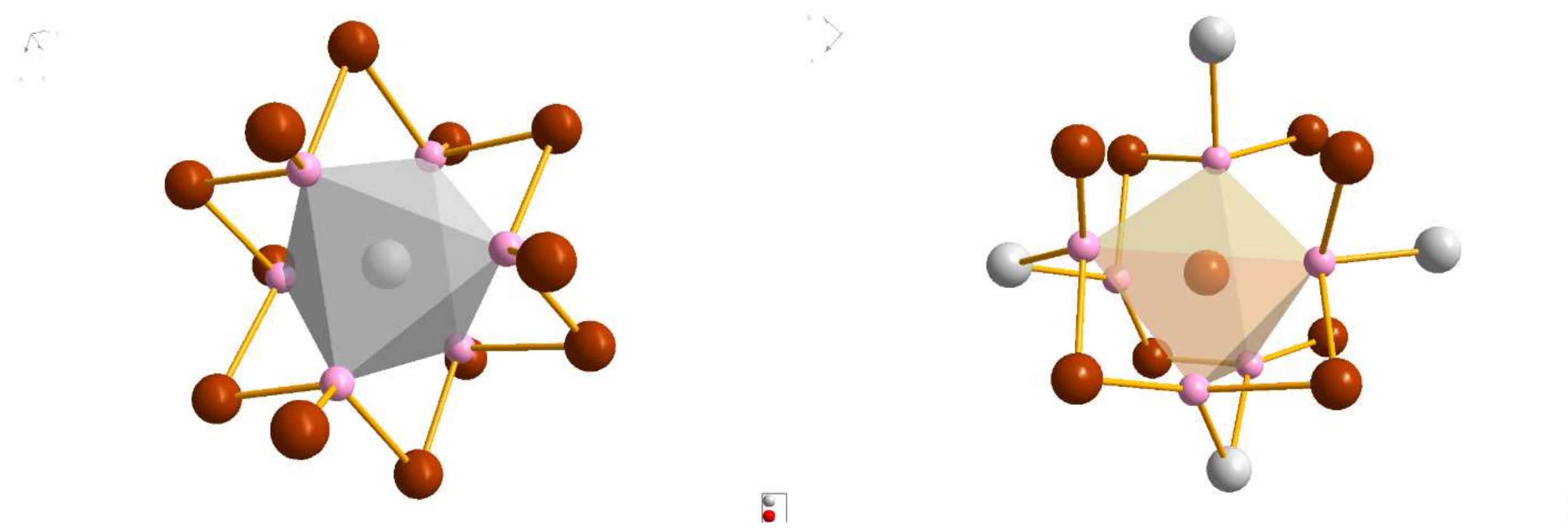

Figure 5. The immediate environment of the cationic positions M1 (grey) and M2 (brown) in $(Mn, Fe, Ga)_2O_3$.

The probability of occurrence of each possible local environment can be calculated using the well-known binomial distribution, defined as follows:

$$P_m^n = \frac{m!}{n!(m-n)!} x^{m-n} (1-x)^n \tag{1}$$

Here, $m$ is the number of metal ions in the nearest coordination shell surrounding the $Fe^{3+}$ ion, $n$ is the number of Mn or Ga ions in a given configuration ($n$ = 0, 1, ..., $m$), and $P_m^n$ is the probability of a configuration containing $n$ Mn/Ga ions and ($m - n$) $Fe^{3+}$ ions. The parameter $x$ denotes the effective iron concentration in the crystal, taking into account site-specific population.

By incorporating the Mössbauer-derived occupancies of the inequivalent crystallographic sites (M1 and M2), the effective site-specific iron concentration ($x$) for each site can be determined for all samples (see Tables 2 and 3). In the bixbyite structure (space group $Ia\bar{3}$.), described in Fig. 1, the second coordination sphere of $Fe^{3+}$ ions in both M1 (8$b$) and M2 (24$d$) sites comprises 12 metal neighbors (Fig. 5).

The calculated probabilities (Eq. 1) are presented in Table 4 and Fig. 6. As can be seen, for sample S1 – with an effective iron concentration of $x = 0.36$ at the M1 (8*b*) site – the probability of a local environment containing $n = 8$ Mn/Ga ions and 4 $Fe^{3+}$ ions (i.e., $m - n = 4$) is approximately 24%.

Table 3. Here, *m* is the number of metal ions in the nearest coordination shell surrounding the $Fe^{3+}$ ion, *n* is the number of Mn or Ga ions in a given configuration ($n = 0, 1, ..., m$), *x* is the iron concentration per formula unit, and *A* is the spectral area, which denotes the relative site occupancy.

| Sample no., space group | *m* | *n* | *x* | *A* |
|---|---|---|---|---|
| S1 $Ia\bar{3}$ | 12 | 0, 1, . . ., 12 | 0.36 | 0.38 |
| | | | 0.58 | 0.62 |
| S2 $Ia\bar{3}$ | 12 | 0, 1, . . ., 12 | 0.28 | 0.36 |
| | | | 0.34 | 0.44 |
| | | | 0.16 | 0.21 |
| S3 $Ia\bar{3}$ | 12 | 0, 1, . . ., 12 | 0.17 | 0.34 |
| | | | 0.12 | 0.24 |
| | | | 0.07 | 0.13 |
| | | | 0.15 | 0.29 |
| S4 $Ia\bar{3}$ | 12 | 0, 1, . . ., 12 | 0.14 | 0.29 |
| | | | 0.10 | 0.20 |
| | | | 0.26 | 0.51 |

Table 4. Probabilities *P* (%) of different local environments for $Fe^{3+}$ ions in (Mn, Fe, $Ga_2O_3$. *x* – iron concentration per formula unit.

| Sample | Position | *x* | 10Fe2MnGa | 9Fe3MnGa | 8Fe4MnGa | 7Fe5MnGa | 6Fe6MnGa | 5Fe7MnGa | 4Fe8MnGa | 3Fe9MnGa | 2Fe10MnGa | 1Fe11MnGa | 12MnGa |
|---|---|---|---|---|---|---|---|---|---|---|---|---|---|
| S1 | M1 | 0.36 | -- | 1 | 2 | 6 | 13 | 21 | 24 | 20 | 10 | 4 | 1 |
| | M2 | 0.58 | 5 | 11 | 19 | 23 | 20 | 12 | 6 | 2 | -- | -- | -- |
| S2 | M1 | 0.28 | -- | -- | 1 | 2 | 6 | 14 | 22 | 25 | 20 | 9 | 2 |
| | M2 | 0.34 | -- | -- | 2 | 5 | 12 | 20 | 23 | 20 | 12 | 4 | 1 |
| | | 0.16 | -- | -- | -- | -- | 1 | 2 | 8 | 18 | 30 | 30 | 13 |
| S3 | M1 | 0.15 | -- | -- | -- | -- | -- | 2 | 6 | 16 | 29 | 31 | 15 |
| | | 0.07 | -- | -- | -- | -- | -- | -- | 1 | 3 | 14 | 37 | 45 |
| | M2 | 0.17 | -- | -- | -- | -- | 1 | 3 | 9 | 20 | 30 | 26 | 11 |
| | | 0.12 | -- | -- | -- | -- | -- | 1 | 4 | 12 | 26 | 35 | 22 |
| S4 | M1 | 0.14 | -- | -- | -- | -- | -- | 2 | 7 | 17 | 29 | 30 | 14 |
| | M2 | 0.26 | -- | -- | -- | 2 | 5 | 13 | 21 | 25 | 21 | 10 | 2 |
| | | 0.10 | -- | -- | -- | -- | -- | -- | 2 | 9 | 24 | 37 | 27 |

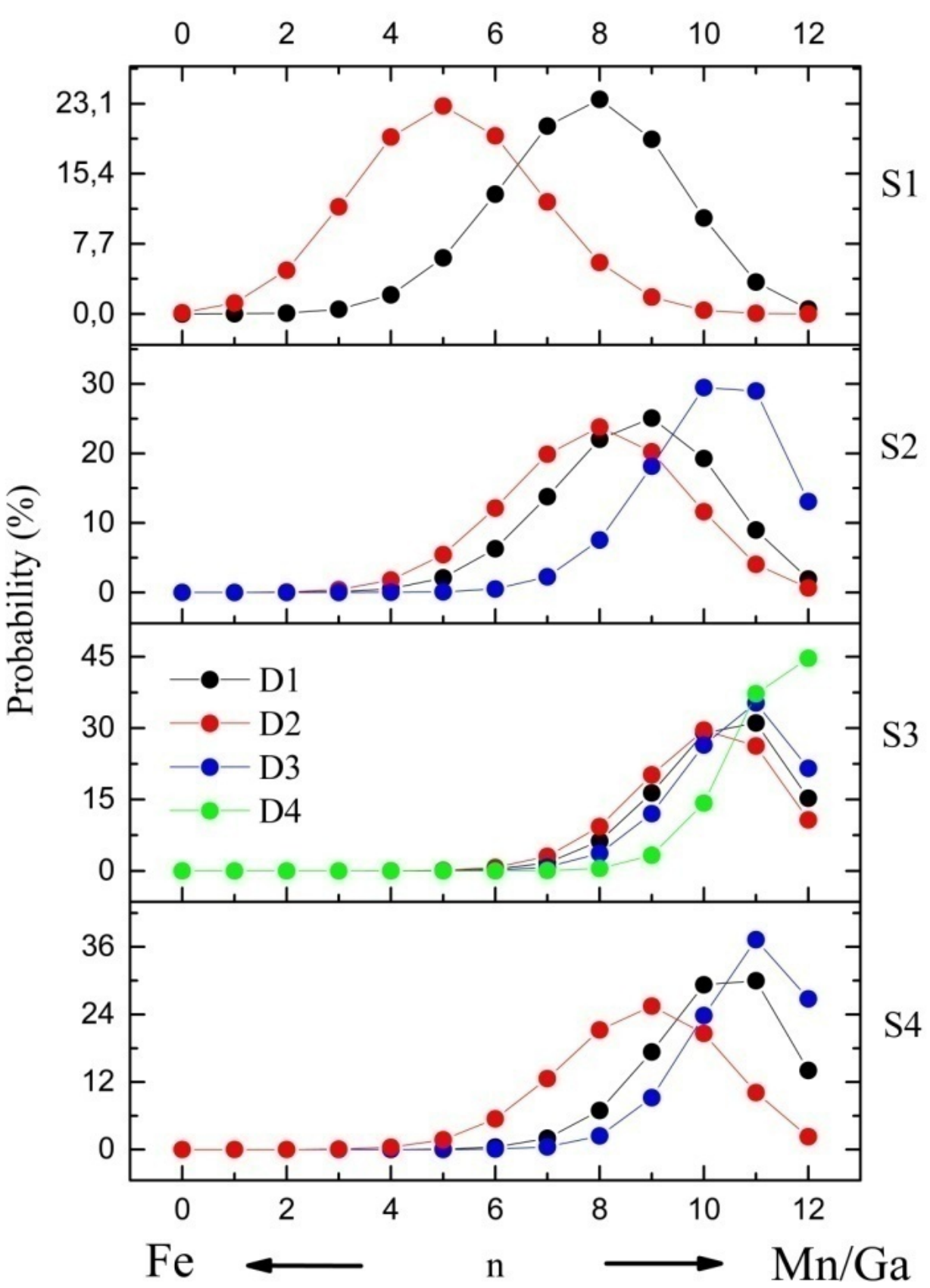

Figure 6. The probability of a configuration containing $n$ manganese or gallium ions and $(m - n)$ iron ions. The colors correspond to the same doublets: black – D1, red – D2, blue – D3, green – D4 (see Table 2).

Sample S1 exhibits the highest degree of cation ordering, as the number of crystallographic sites corresponds exactly to the number of doublets obtained from the analysis of Mössbauer spectra (see Table 2). The most probable configurations of the twelve metal ions surrounding $Fe^{3+}$ are as follows: for site M2 (24*d*): $9Fe^{3+}$ – 3(Mn, Ga), $8Fe^{3+}$ – 4(Mn, Ga), $7Fe^{3+}$ – 5(Mn, Ga), $6Fe^{3+}$ – 6(Mn, Ga), $5Fe^{3+}$ – 7(Mn, Ga); for site M1 (8*b*): $6Fe^{3+}$ – 6(Mn, Ga), $5Fe^{3+}$ – 7(Mn, Ga), $4Fe^{3+}$ – 8(Mn, Ga), $3Fe^{3+}$ – 9(Mn, Ga), $2Fe^{3+}$ – 10(Mn, Ga).

Sample S2 shows a lower degree of cation ordering, as an additional inequivalent state appears within the M2 site. The most probable $Fe^{3+}$ environments are: for site M1 (8*b*): $5Fe^{3+}$ – 7(Mn, Ga), $4Fe^{3+}$ – 8(Mn, Ga), $3Fe^{3+}$ – 9(Mn, Ga), $2Fe^{3+}$ – 10(Mn, Ga), $1Fe^{3+}$ – 11(Mn, Ga); for site M2 (24*d*) at $x = 0.34$: $6Fe^{3+}$ – 6(Mn, Ga), $5Fe^{3+}$ – 7(Mn, Ga), $4Fe^{3+}$ – 8(Mn, Ga), $3Fe^{3+}$ – 9(Mn, Ga), $2Fe^{3+}$ – 10(Mn, Ga); for site M2 (24*d*) at $x = 0.16$: $3Fe^{3+}$ – 9(Mn, Ga), $2Fe^{3+}$ – 10(Mn, Ga), $1Fe^{3+}$ – 11(Mn, Ga), $0Fe^{3+}$ – 12(Mn, Ga).

Sample S3 exhibits an even lower degree of cation ordering. Two crystallographic sites give rise to four inequivalent $Fe^{3+}$ environments with distinct local distortions (see Table 2). The most probable configurations for $Fe^{3+}$ at site M1 (8*b*) are: $3Fe^{3+}$ – 9(Mn, Ga), $2Fe^{3+}$ – 10(Mn, Ga), $1Fe^{3+}$ – 11(Mn, Ga), $0Fe^{3+}$ – 12(Mn, Ga). The emergence of multiple states within a single crystallographic site indicates a non-uniform distribution of iron cations over the lattice positions. The M2 (24*d*) site in this sample splits into two inequivalent states with different local environments and site occupancies (Tables 3, 4). The most probable $Fe^{3+}$ configurations for M2, along with their relative abundances accounting for site population, are listed in Table 4.

Sample S4 shows reduced cation ordering compared to S1. Two crystallographic sites correspond to three inequivalent $Fe^{3+}$ states with different local distortions (see Table 2). For site M1 (8*b*), the most probable $Fe^{3+}$ environments are: $3Fe^{3+}$ – 9(Mn, Ga), $2Fe^{3+}$ – 10(Mn, Ga), $1Fe^{3+}$ – 11(Mn, Ga), $0Fe^{3+}$ – 12(Mn, Ga). The M2 (24*d*) site in this sample also splits into two inequivalent positions with distinct local coordination and occupancies (Tables 3 and 4).

*3.2.2. X-ray Absorption Spectroscopy*

XANES spectra of the ternary oxides $(Mn, Fe, Ga)_2O_3$ at the K-edges of Mn, Fe, and Ga, together with the spectra of standards, are shown in Fig. 7. The spectra of the studied samples exhibit only minor differences at all three X-ray absorption edges, indicating that the basic local structure is similar across all samples. However, variations in interatomic distances or site occupancies may still be present. Based on the position of the absorption edge, which coincides with that of the reference compound $Fe_2O_3$, the oxidation state of iron in all samples is +3. Similarly, the edge position in the Mn K-edge XANES spectra indicates that manganese also has a +3 oxidation state in all samples, consistent with iron. The intense "white line" and the shift of the absorption edge to higher energies relative to metallic Ga further confirm that gallium is in an oxidized state, consistent with $Ga^{3+}$.

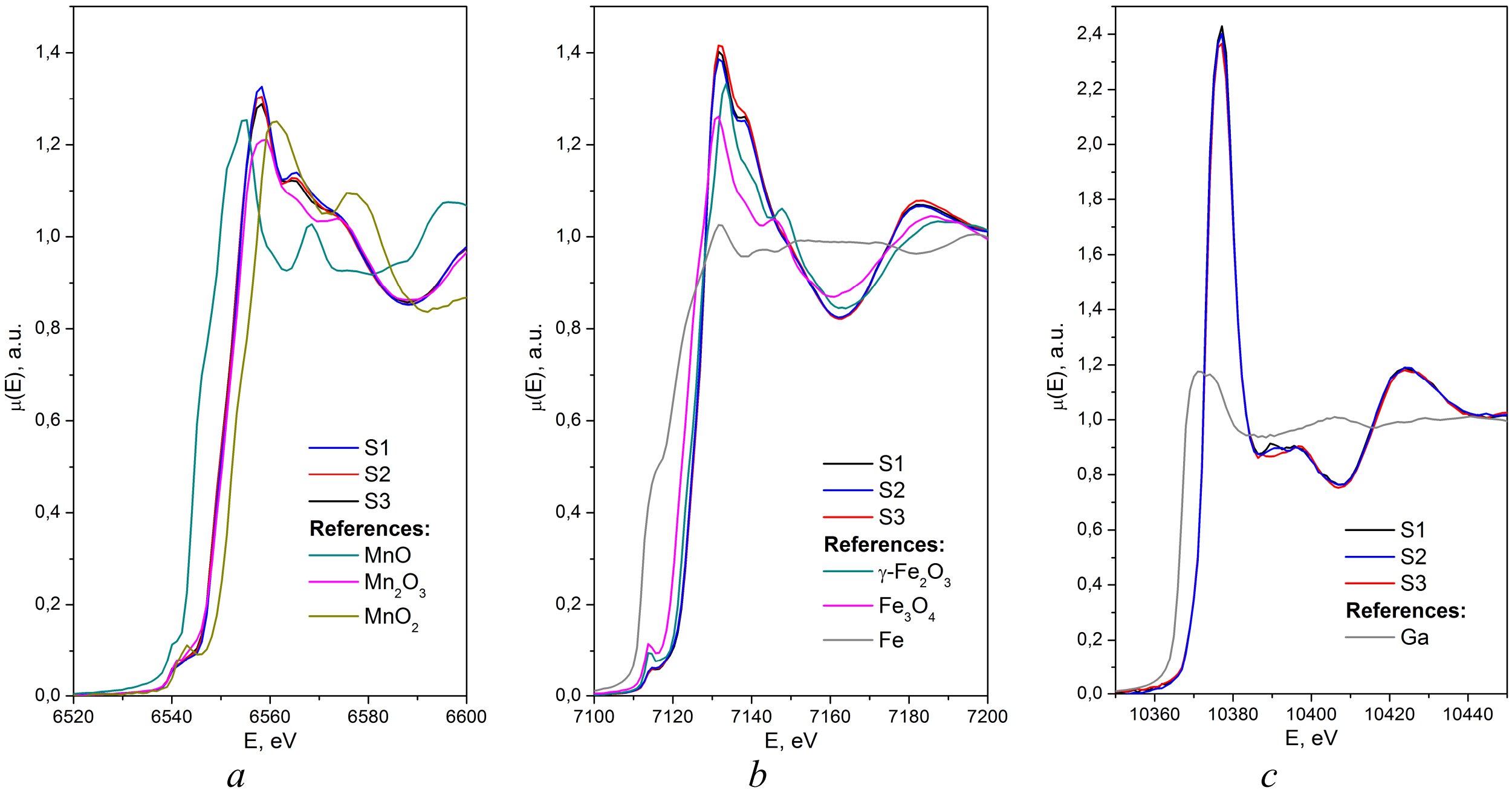

Figure 7. XANES spectra of the $(Mn, Fe, Ga)_2O_3$ triple oxides (S1, S2 and S3 samples) at the Mn (a), Fe (b) and Ga (c) K-edges. The spectra are presented along with the reference material spectra (MnO, $Mn_2O_3$ and $MnO_2$ for the Mn K-edge; $\gamma$-$Fe_2O_3$, $Fe_3O_4$ and the Fe for the Fe K-edge; Ga for the Ga K-edge).

The Fourier transform of the Fe K-edge EXAFS spectra for all three samples exhibits four well-resolved peaks, corresponding to the coordination shells surrounding Fe atoms (Fig. 8). The first shell ($R$ = 1.5-2.0 Å) is attributed to oxygen neighbors (Fe-O), while the second, corresponding to the most intense peak in the transform, arises from metal atoms (Fe-M, where M = Mn, Fe, Ga). Differences in peak intensities among the three samples are minor—no more than 10–15% – which is within the experimental uncertainty in coordination number (CN) determination.

The Fourier transforms of the Ga K-edge EXAFS spectra also show four peaks, similar to those observed at the Fe K-edge (Fig. 8). For samples S1 and S3, the local environment around gallium is nearly identical, as evidenced by the same peak positions and intensities. In contrast, sample S2 shows a significant increase in the intensity of the first two shells and a shift of the third shell peak toward larger interatomic distances.

The Mn K-edge EXAFS Fourier transforms are similar to those observed for Fe and Ga. Based on the positions and intensities of the peaks, the local coordination environment of manganese is nearly the same across all three samples. Notably, in contrast to the Fe and Ga edges, no shift in the third coordination shell peak is observed as a function of gallium concentration, indicating a more uniform Mn local structure.

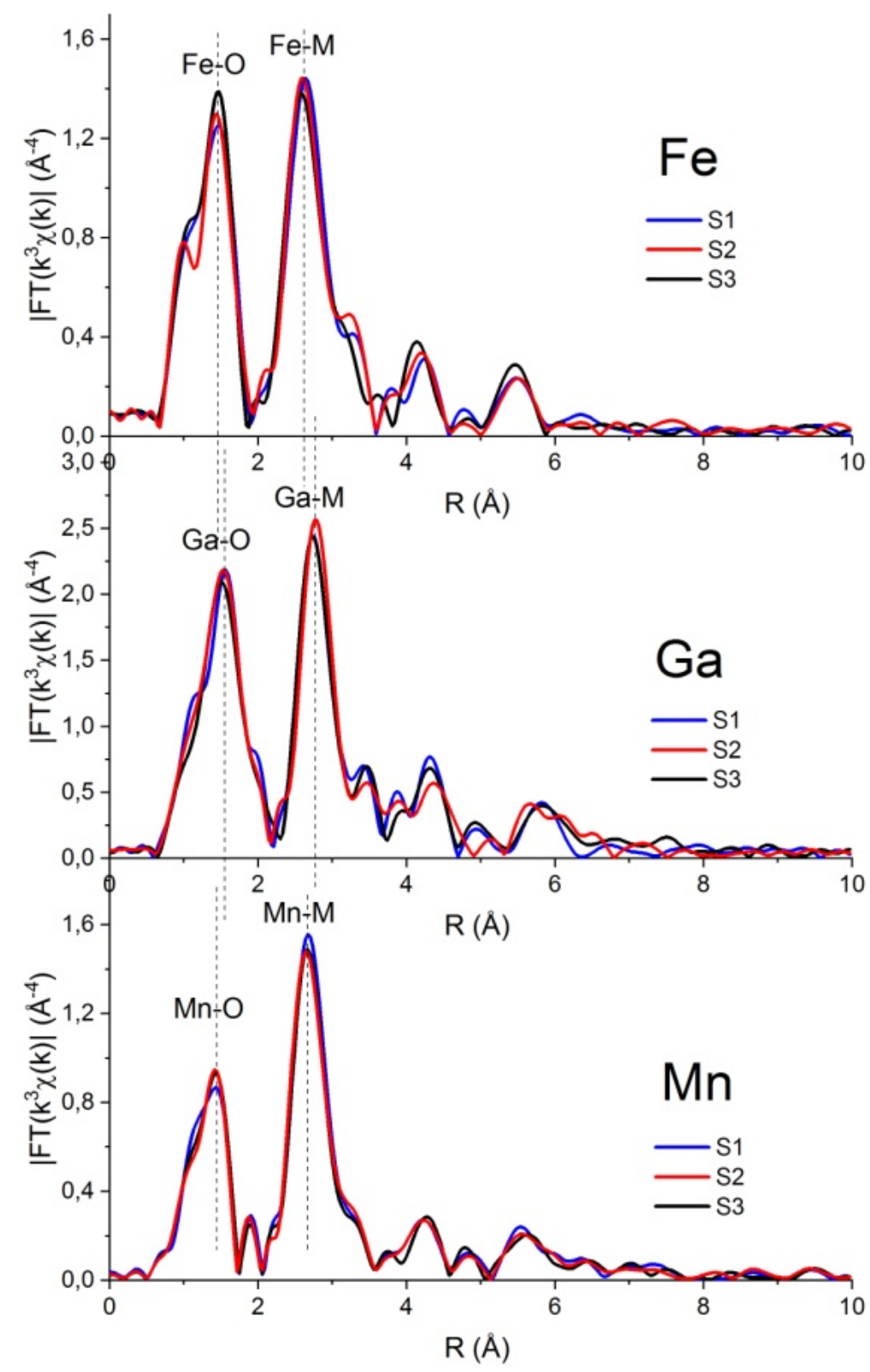

Figure 8. EXAFS spectra of the (Mn, Fe, Ga)$_2$O$_3$ triple oxides (samples S1, S2 and S3).

Quantitative analysis and refinement of the local structure of (Mn, Fe, Ga)$_2$O$_3$ were performed through EXAFS spectral modeling. Due to the large number of possible cation configurations – arising from the presence of three different cations and two crystallographic sites, one of which is non-centrosymmetric – determining the precise coordination environment in these systems is a challenging task. Therefore, a simplified fitting model was employed to determine the bond lengths, in which each element was assumed to be surrounded exclusively by manganese cations in the second coordination shell. The resulting structural parameters are summarized in Table 5. The complete fitting results, including coordination numbers, bond distances, and Debye-Waller factors, are provided in Table S1.

Table 5. Local structure (coordination shell radii, in Å) determined by EXAFS of each the cation type (Fe, Mn, Ga) in (Mn, Fe, Ga)$_2$O$_3$ triple oxides (S1, S2 and S3 samples). $R_f$ is the numerical value characterizing the difference between the fitting curve and the experimental data. M1 is 8*b* position, M2 is 24*d* position. O1_1 (blue) is oxygen in the first coordination sphere of 8*b*. O1_1, O1_2, O1_3 (light-red) are oxygen in the first coordination sphere of 24*d*. Purple denotes the second coordination sphere of M1 (8*b*) and M2 (24*d*). Mn1_1 and Mn1_2 – cations in M1 position at different distances from M; Mn2_1 and Mn2_2 – cations in M2 position at different distances from M.

| Fe | | | |
|---|---|---|---|
| Scattering path | S1 $Fe_{0.93}Mn_{0.68}Ga_{0.39}O_3$ | S2 $Fe_{0.78}Mn_{0.68}Ga_{0.54}O_3$ | S3 $Fe_{0.50}Mn_{0.84}Ga_{0.66}O_3$ |
| $R_f$, % | 1.4 | 1.4 | 1.3 |
| Fe1 – O1_1 | 2.03 ± 0.02 | 2.03 ± 0.02 | 1.99 ± 0.02 |
| Fe1 – Mn2_1 | 3.11 ± 0.01 | 3.10 ± 0.01 | 3.08 ± 0.02 |
| Fe1 – Mn2_2 | 3.57 ± 0.03 | 3.59 ± 0.02 | 3.54 ± 0.04 |
| Fe2 – O1_1 | 1.94 ±0.02 | 1.94 ±0.02 | 1.90 ±0.02 |
| Fe2 – O1_2 | 1.95 ±0.11 | 1.96 ±0.10 | 1.84 ±0.59 |
| Fe2 – O1_3 | 2.12 ±0.11 | 2.13 ±0.10 | 2.01 ±0.59 |
| Fe2 – Mn1_1 | 3.11 ±0.01 | 3.10 ±0.01 | 3.08 ±0.02 |
| Fe2 – Mn2_1 | 3.13 ±0.01 | 3.12 ±0.02 | 3.10 ±0.02 |
| Fe2 – Mn1_2 | 3.57 ±0.03 | 3.59 ±0.02 | 3.54 ±0.04 |
| Fe2 – Mn2_2 | 3.73 ±0.16 | 4.05 ±0.10 | 3.23 ±0.75 |
| **Mn** | | | |
| Scattering path | S1 $Fe_{0.93}Mn_{0.68}Ga_{0.39}O_3$ | S2 $Fe_{0.78}Mn_{0.68}Ga_{0.54}O_3$ | S3 $Fe_{0.50}Mn_{0.84}Ga_{0.66}O_3$ |
| $R_f$, % | 1.3 | 1.3 | 1.2 |
| Mn1 – O1_1 | 1.93 ± 0.02 | 1.91 ± 0.02 | 1.92 ± 0.02 |
| Mn1 – Mn2_1 | 3.08 ± 0.01 | 3.06 ± 0.01 | 3.06 ± 0.03 |
| Mn1 – Mn2_2 | 3.48 ± 0.03 | 3.50 ± 0.03 | 3.55 ± 0.06 |
| Mn2 – O1_1 | 1.80 ±0.04 | 1.80 ±0.06 | 1.85 ±0.08 |
| Mn2 – O1_2 | 2.08 ±0.06 | 2.07 ±0.09 | 2.03 ±0.15 |
| Mn2 – O1_3 | 2.21 ±0.06 | 2.18 ±0.05 | 2.17 ±0.09 |
| Mn2 – Mn1_1 | 3.08 ±0.01 | 3.06 ±0.01 | 3.06 ±0.03 |
| Mn2 – Mn2_1 | 3.20 ±0.06 | 3.17 ±0.06 | 3.15 ±1.61 |
| Mn2 – Mn1_2 | 3.48 ±0.03 | 3.50 ±0.03 | 3.55 ±0.06 |
| Mn2 – Mn2_2 | 3.64 ±0.04 | 3.68 ±0.07 | 3.15 ±1.58 |
| **Ga** | | | |
| Scattering path | S1 $Fe_{0.93}Mn_{0.68}Ga_{0.39}O_3$ | S2 $Fe_{0.78}Mn_{0.68}Ga_{0.54}O_3$ | S3 $Fe_{0.50}Mn_{0.84}Ga_{0.66}O_3$ |
| $R_f$, % | 1.4 | 1.2 | 1.2 |
| Ga1 – O1_1 | 2.01 ± 0.03 | 1.99 ± 0.02 | 2.01 ± 0.03 |
| Ga1 – Mn2_1 | 3.11 ± 0.02 | 3.15 ± 0.01 | 3.11 ± 0.02 |
| Ga1 – Mn2_2 | 3.57 ± 0.12 | 3.60 ± 0.03 | 3.56 ± 0.10 |
| Ga2 – O1_1 | 1.90 ±0.10 | 2.13 ±1.85 | 1.88 ±3.06 |
| Ga2 – O1_2 | 2.14 ±0.13 | 1.50 ±0.03 | 1.53 ±0.06 |
| Ga2 – O1_3 | 2.22 ±0.10 | 2.13 ±1.81 | 1.88 ±3.01 |
| Ga2 – Mn1_1 | 3.12 ±0.02 | 3.15 ±0.01 | 3.11 ±0.02 |
| Ga2 – Mn2_1 | 3.28 ±0.14 | 3.16 ±0.01 | 3.26 ±0.17 |
| Ga2 – Mn1_2 | 3.57 ±0.12 | 3.60 ±0.03 | 3.56 ±0.10 |
| Ga2 – Mn2_2 | 3.69 ±0.62 | 3.74 ±0.08 | 3.65 ±0.55 |

The applied fitting model yielded good agreement between the experimental and calculated EXAFS spectra for Fe and Mn within the first two coordination shells in all three samples. For Ga, however, only a moderate agreement was achieved, and only for sample S1, which has the lowest Ga content. Below follows a systematic analysis of the local environments at the centrosymmetric M1 (8*b*) and non-centrosymmetric M2 (24*d*) sites, focusing first on the anion (oxygen) and then on the cation (metal) coordination shells.

In the first coordination sphere, the Fe1-O and Ga1-O bond distances were found to be in the range of 1.99-2.03 Å, with uncertainties below 2%. These values are in good agreement with previously reported bond lengths in bixbyites [25]. In contrast, the Mn1-O distances are significantly shorter across all samples, ranging from 1.91 to 1.93 Å – despite the fact that the ionic radii of $Mn^{3+}$ and $Fe^{3+}$ are equal (0.645 Å), and $Ga^{3+}$ has an even smaller ionic radius (0.62 Å). This discrepancy suggests that Mn is unlikely to occupy the centrosymmetric M1 (8*b*) site. Instead, the observed Mn-O bond length of 1.91-1.93 Å is more consistent with the distorted environment of the M2 (24*d*) site. Analysis of the anion coordination around M2 further supports this interpretation. For all three samples (S1, S2, and S3), a good fit between the experimental and modeled EXAFS data was obtained only for Mn at the M2 site. Three distinct Mn2-O bond distances were determined with uncertainties up to 4%: 1.80-1.85 Å, 2.03-2.08 Å, and 2.17-2.21 Å. Within experimental error, these values are nearly identical across the samples (Table 5). This type of asymmetric oxygen coordination is characteristic of Mn-rich bixbyites [25].

The anion environments of Fe2 and Ga2 in sample S1 (which has the highest Fe and lowest Ga content) differ slightly from that of Mn, but overall resemble an averaged coordination derived from single-crystal X-ray diffraction – where the shortest bonds are elongated to minimize octahedral distortion. For comparison, the corresponding bond lengths in the Mn-rich bixbyite $Fe_{0.034}Mn_{1.966}O_3$ are 1.8975 Å, 1.9870 Å, and 2.2423 Å [28]. In sample S1, single-crystal XRD yields M2-O distances of 1.9305 Å, 2.0341 Å, and 2.1399 Å [25]. A comparison of these XRD-derived bond lengths with XAS results for Mn2-O and Fe2-O indicates that both Fe and Mn cations occupy the distorted M2 (24*d*) site. In contrast, the corresponding Ga bond lengths (1.90 Å, 2.146 Å, 2.22 Å for S1 – the most reliable among the three samples) deviate more significantly from the XRD data, suggesting a less favorable fit for Ga in this model.

As the Fe concentration decreases (in samples S2 and especially S3), the EXAFS fitting results for Fe and Ga become increasingly uncertain and cannot be reliably interpreted.

Analysis of the cationic second coordination shells (M1-M and M2-M) shows relatively small uncertainties for Fe and Mn. These results are consistent with Mössbauer spectral analysis based on a binomial distribution model, which indicates that Fe ions are predominantly surrounded by non-Fe cations. This qualitatively supports the simplified fitting assumption that the second shell around Fe consists entirely of $Mn^{3+}$ ions.

A comparative analysis of the M1–M and M2–M bond distances reveals only minor differences between the samples, within the experimental uncertainties.

Assuming that Ga preferentially occupies the centrosymmetric M1 (8*b*) site, and incorporating the Fe site distribution derived from Mössbauer spectroscopy, the cation distributions for the samples can be expressed as:

- **S1**: $(Fe_{0.35}Ga_{0.15})^1(Fe_{0.58}Ga_{0.24}Mn_{0.68})^2O_3$
- **S2**: $(Fe_{0.28}Ga_{0.22})^1(Fe_{0.51}Ga_{0.32}Mn_{0.68})^2O_3$
- **S3**: $(Fe_{0.21}Ga_{0.29})^1(Fe_{0.29}Ga_{0.37}Mn_{0.84})^2O_3$
- **S4** (from Mössbauer effect): $(Fe_{0.15}Mn_{0.35})^1(Fe_{0.37}Mn_{1.13})^2O_3$

Each of the studied bixbyite samples exhibits significant **positional cation disorder**, with Fe and Ga distributed across both crystallographic sites, while Mn preferentially occupies the distorted M2 (24*d*) site.

## 4. Discussion

Thus, we are dealing with four (Fe, Mn, Ga)$_2$O$_3$ samples exhibiting different Fe/Mn/Ga cation ratios, which display similar structural and magnetic properties overall, yet also show certain inconsistencies. According to magnetometry data, a magnetic phase transition is observed in sample S2 at $T \approx 370$ K, tentatively attributed to a ferrimagnetic ordering – similar to that reported for some bixbyite-type oxides in the literature. No such feature is observed in the other three samples, including the gallium-free one.

The low-temperature magnetization behavior of all four samples is qualitatively similar: low-temperature anomalies are present in the range $T \approx 20$-35 K, depending on the sample. Magnetization measurements were performed on single-crystalline specimens – individual crystals up to $2\times2\times2$ mm$^3$ in size, approximately cubic in shape. However, Mössbauer spectroscopy results appear to contradict these findings: at room temperature, all four samples exhibit doublet spectra, indicating that iron cations are in a paramagnetic state. The Mössbauer measurements were carried out on powdered samples prepared from carefully selected single crystals. Since powder spectra represent an ensemble average over many crystallites, this technique provides a more statistically representative and integrated characterization compared to single-crystal magnetometry. Therefore, the central question of this study arises: what is the origin of the apparent ferrimagnetic behavior in the crystal $Fe_{0.78}Mn_{0.68}Ga_{0.54}O_3$ (sample S2)?

Analysis of the cation distribution over the two inequivalent crystallographic sites (M1: 8*b* and M2: 24*d*) reveals that none of the samples exhibit complete cation ordering, with no exclusive site preference for any single cation type. However, certain trends in cation site occupancy are evident and depend on the Fe/Mn/Ga ratio. In particular, Mn shows a clear preference for the distorted M2 (24*d*) site.

The distribution of Fe over the two sites varies irregularly across the series: $Fe_1/Fe_2 = 0.6$ (S1), 0.55 (S2), 0.72 (S3), and 0.41 (S4). Notably, even in sample S4 – where Ga is absent and thus does not compete for cation sites – Fe still occupies both M1 and M2 positions, indicating that site occupancy is not solely governed by chemical competition.

The local environment in the second coordination shell and the higher number of Fe sites in the $(Fe, Mn, Ga)_2O_3$ solid solutions compared to the nominal crystallographic sites may be related to differences in crystal growth kinetics. Sample S1 was grown with a slow cooling rate of $dT/dt = 4$ °C/day, S2 at a faster rate of 8 °C/day, and S3 at 12 °C/day. In contrast, the Ga-free sample S4 was synthesized under isothermal conditions but from a significantly different flux composition, which likely resulted in altered diffusion kinetics. This suggests that diffusion rates during crystal growth play a key role in determining the degree of cation site ordering in these oxides.

To gain deeper insight into the temperature ranges of the observed magnetic anomalies, the temperature derivatives $\partial(\chi\cdot T)/\partial T(T)$ [29-30] were analyzed in the low-temperature region (Fig. 9a), and the derivative $\partial(M^2)/\partial T(T)$ [31] was examined in the vicinity of the ferrimagnetic phase transition for sample S2 (Fig. 9b).

Qualitatively, the low-temperature behavior of $\partial(\chi\cdot T)/\partial T(T)$ is similar across all three samples. A broad, asymmetric maximum is observed in the temperature range 10-40 K. With sufficient resolution provided by fine temperature steps, the data for samples S1 and S3 reveal that this maximum consists of two overlapping peaks, indicating the presence of two successive phase transitions. The shape of the maximum in sample S2 is analogous to that in S1 and S3, suggesting a similar two-step transition process.

As the iron content decreases, the transition temperatures shift to lower values. Comparing these results with the data in Fig. 2, the lower-temperature peak coincides with the bifurcation temperature of the FC and ZFC curves. This low-temperature anomaly can thus be associated with a spin-glass-type phase transition, consistent with earlier interpretations based on AC magnetic susceptibility measurements of samples S1 and S3 [25].

In the temperature dependences of $\chi'$ and $\chi''$, as well as in the magnetization and its derivatives, two distinct anomalies are observed. The higher-temperature anomaly is accompanied by a relatively sharp peak in magnetic susceptibility, while the low-temperature feature is characterized by a broad, weak maximum that exhibits significantly stronger frequency dependence than the high-temperature one. Analysis of the Mydosh parameter confirms that the frequency dependence is more pronounced in sample S1 than in S3. Sample S1 was previously classified as a canonical spin glass below the higher-temperature transition $T_1$, whereas in sample S3, this anomaly does not correspond to conventional spin freezing. These differences in low-temperature magnetic behavior between samples S1 and S3 may originate from variations in the degree of local cationic ordering, as inferred from the binomial distribution analysis of site occupancies.

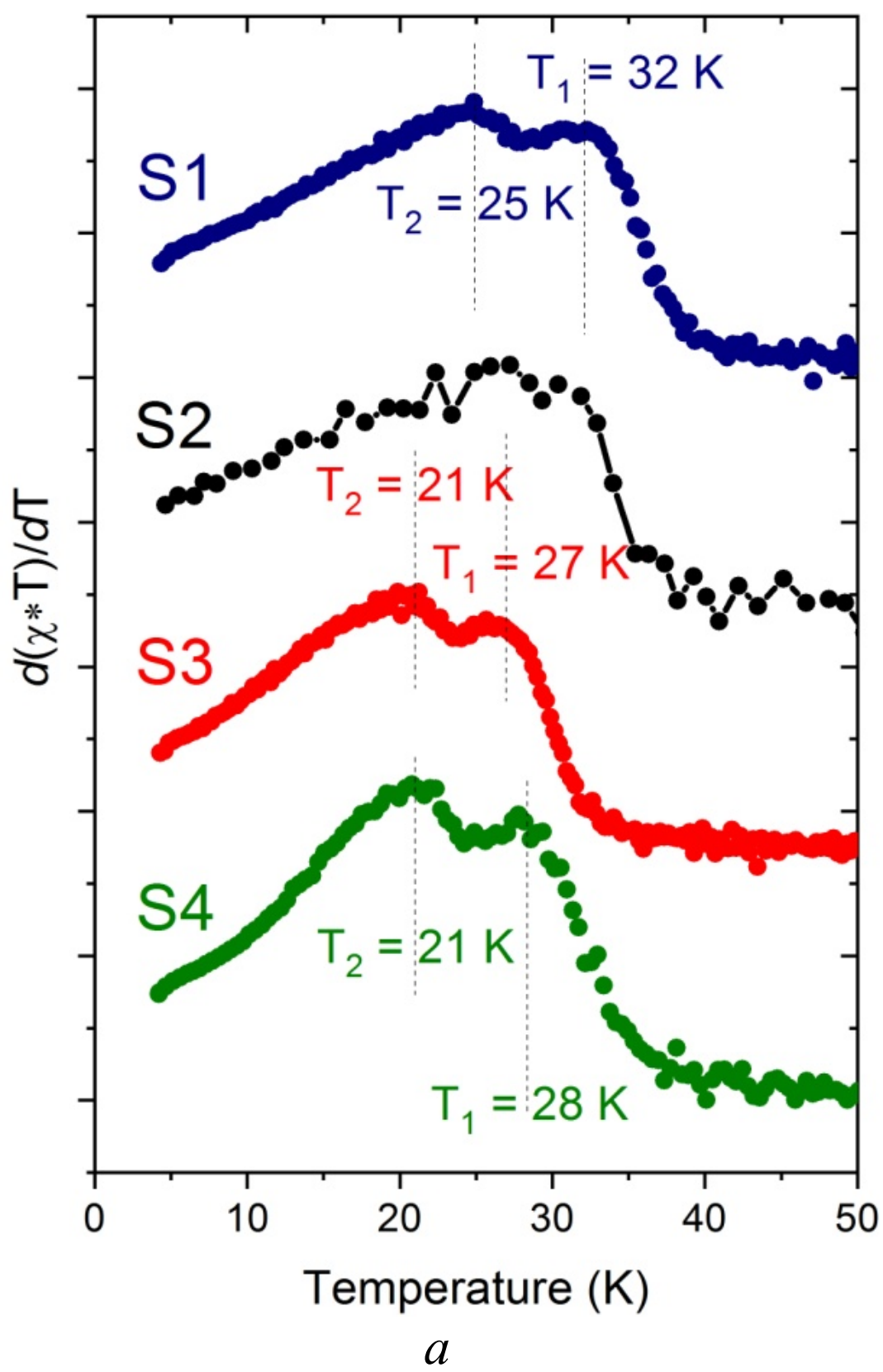

*a*

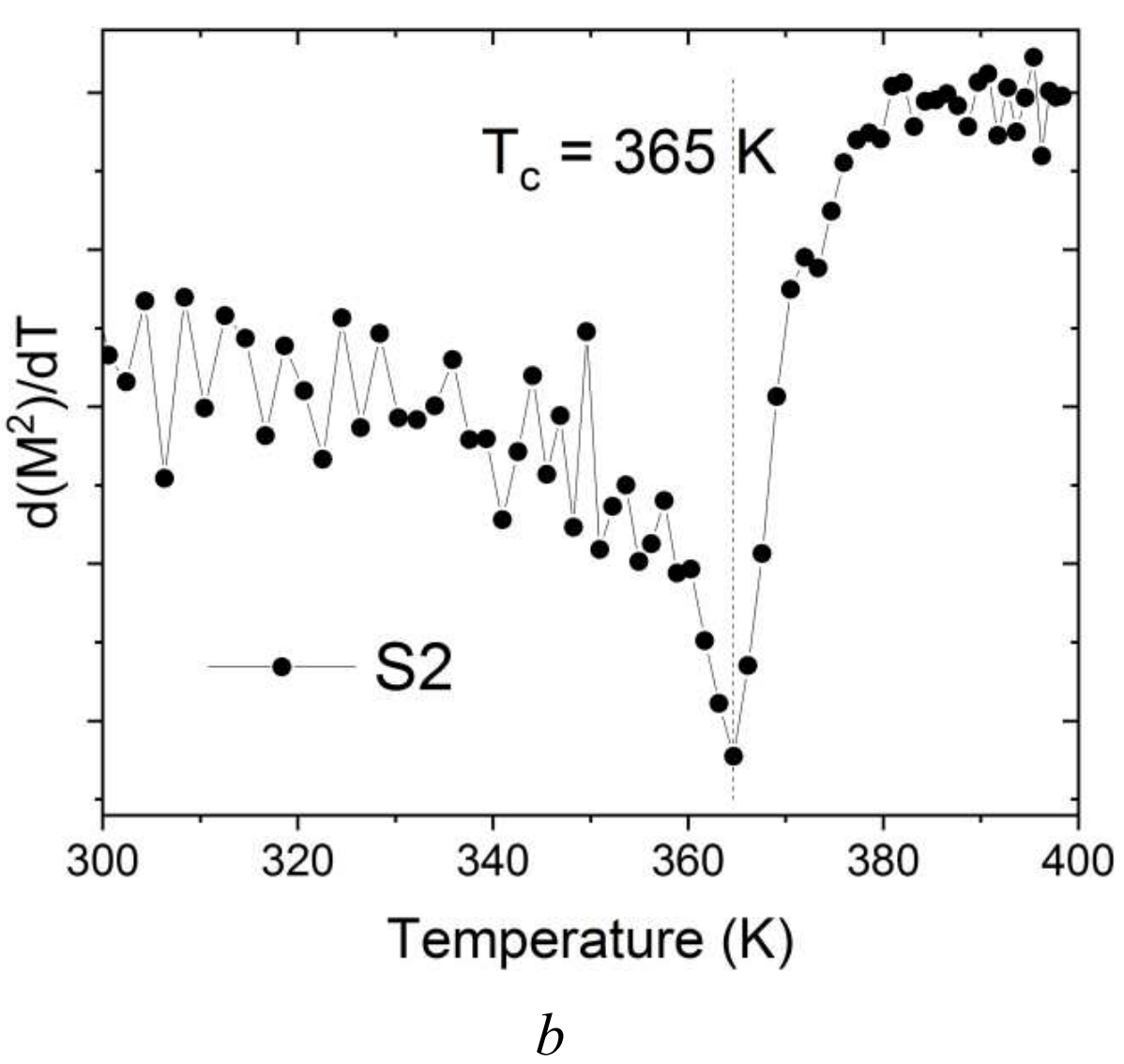

*b*

Figure 9. Thermal dependences $\partial(\chi\cdot T)/\partial T(T)$ (*a*) and $\partial(M^2)/\partial T(T)$ (*b*) for S1-S3 and S2 samples, respectively. The plots illustrate the actual phase transition temperatures of the low-temperature antiferromagnetic (or spin-glass) and high-temperature ferrimagnetic phase transitions.

To gain further insight into the low-temperature anomalies, temperature-dependent specific heat measurements were performed on sets of single crystals of samples S1-S4 in the temperature range $\Delta T$ = 7-60 K (Fig. 10). Broad, weak anomalies in the heat capacity were observed in all samples within the temperature range 15-30 K.

The experimental data were analyzed by fitting the temperature dependencies above 40 K using a linear combination of Debye and Einstein models, both with and without an additional linear term accounting for possible contributions from cationic and magnetic disorder. Due to the small amplitudes of the observed heat capacity anomalies, it is difficult to unambiguously confirm the presence of a true thermodynamic phase transition.

However, comparison of different fitting approaches reveals that including the linear term leads to transition temperatures that correlate significantly better with those obtained from ac and dc magnetization measurements, as well as from the $\partial(\chi\cdot T)/\partial T(T)$ derivative, which is proportional to the magnetic contribution to the heat capacity in simple antiferromagnets (Figs. 2 and 9). The excess heat capacity (after subtracting the phonon contribution) is shown in the inset of Fig. 10.

Objectively determining the number and precise temperatures of the phase transitions is not feasible due to the broadness and low amplitude of the features – the fitted peak positions strongly depend on the selected temperature range. Nevertheless, based on the shape of the anomalies (Fig. 10), it can be concluded that the observed transitions are not characteristic of antiferromagnetic ordering. In the case of antiferromagnetic transitions, an asymmetric $\lambda$-shaped peak is typically observed – even in cation-disordered solid solutions [32]. Instead, the observed features are more consistent with those expected in magnetically disordered systems, such as spin glasses.

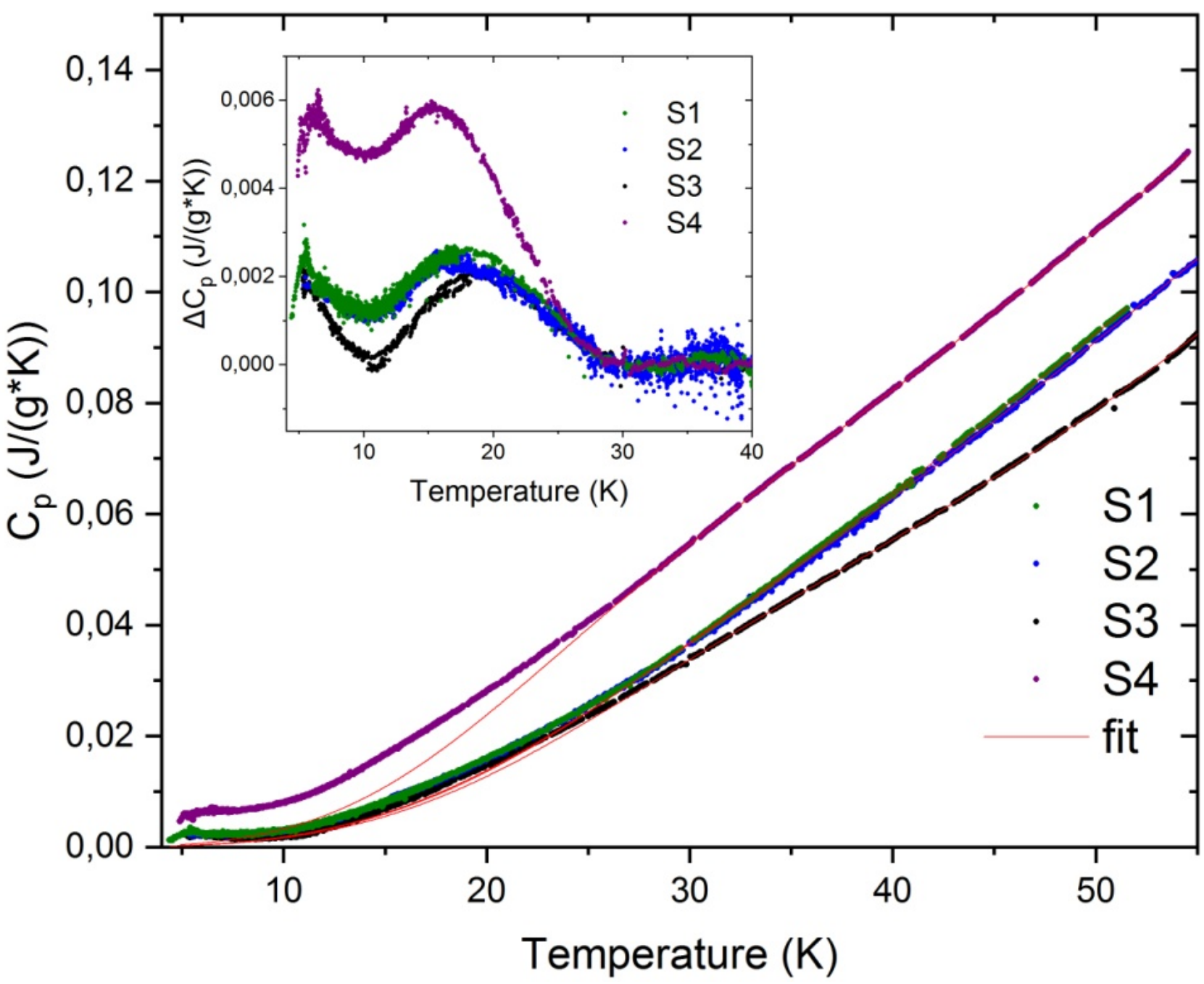

Figure 10. Thermal dependences of specific heat of S1 (green), S2 (blue), S3 (black), S4 (purple) samples and Debye-Einstein fit (red line). Inset: thermal dependences of excess specific heat of S1-S4 samples.

A similar specific heat temperature dependence has been previously reported for the bixbyite $Fe_{1.1}Mn_{0.9}O_3$, which enters a spin-glass state below 32 K [33]. The authors of Ref. [34] suggest that all mixed bixbyites lack long-range magnetic order, and that the magnetic phase transition in these systems corresponds to a transition into a spin-glass state driven by cationic disorder. This hypothesis was experimentally confirmed for $Fe_{1.12}Mn_{0.88}O_3$ through neutron diffraction and inelastic neutron scattering studies, which revealed a correlation length of only 15 Å [34]. In contrast, in undoped $Mn_2O_3$ – which exhibits long-range antiferromagnetic order [35] – a sharp, intense $\lambda$-shaped anomaly in the heat capacity is observed at $T_N$ = 79.5 K, characteristic of a conventional antiferromagnetic phase transition [36].

Analysis of the local cationic environment, based on binomial distribution modeling and supported by the number of components observed in the Mössbauer spectra, indicates that sample S3 exhibits the highest degree of cationic disorder among the ternary oxides. This is consistent with the higher cooling rates used during crystal growth from the flux. In contrast, sample S1 is significantly more ordered. The distribution analysis shows that Fe cations in S1 are surrounded predominantly by other Fe ions, whereas in S3, the local environment is enriched in Mn and Ga. Such a configuration likely weakens magnetic exchange pathways in S3, which is consistent with the lower transition temperature observed in this sample.

The magnetic behavior of the studied samples was analyzed by fitting the temperature dependence of the inverse molar susceptibility in the paramagnetic region using the modified Curie–Weiss law [37] (Fig. 11). The fitting procedure followed the approach described in Ref. [38]. To reduce the number of free parameters and improve the reliability of the fits, the temperature-independent contribution $\chi_0$ – comprising the diamagnetic susceptibility and Van Vleck paramagnetism – was estimated independently (Table 6). The results of the Curie-Weiss fits to the inverse magnetic susceptibility in the high-temperature paramagnetic region, well above the magnetic transition temperatures, are summarized in Table 7.

Table 6. Molar masses $M$ and estimated temperature-independent contribution to magnetic susceptibility of S1 and S3 samples: diamagnetic susceptibility ($\chi_d$), paramagnetic Van Vleck susceptibility ($\chi_{VV}$) and its sum ($\chi_0$).

| Compound | $M$ (g/mol) | $\chi_d \cdot 10^{-4}$ (emu/(mol·Oe)) | $\chi_{VV} \cdot 10^{-4}$ (emu/(mol·Oe)) | $\chi_0 \cdot 10^{-4}$ (emu/(mol·Oe)) |
|---|---|---|---|---|
| $Fe_{0.93}Mn_{0.68}Ga_{0.39}O_3$ (S1) | 164.487 | -0.5522 | 0.2788 | -0.2734 |
| $Fe_{0.78}Mn_{0.68}Ga_{0.54}O_3$ (S2) | 166.569 | -0.5492 | 0.2788 | -0.2704 |
| $Fe_{0.5}Mn_{0.84}Ga_{0.66}O_3$ (S3) | 168.089 | -0.5468 | 0.3444 | -0.2024 |
| $Fe_{0.52}Mn_{1.48}O_3$ (S4) | 158.3484 | -0.5600 | 0.6068 | 0.0468 |

The best agreement between the Curie constants and effective magnetic moments obtained from fitting and those estimated theoretically using the formula $\mu_{eff}^{theor} = \mu_B\left(n(Fe^{3+})g_S^2(Fe^{3+})S(S+1) + n(Mn^{3+})g_S^2(Mn^{3+})S(S+1)\right)^{1/2}$ (where $S$ ($Fe^{3+}$) = 5/2; $S$ ($Mn^{3+}$) = 2; $g_S$ ($Fe^{3+}$) = 2 [39]; $g_S$ ($Mn^{3+}$) = 2 [40]) is observed for sample S3.

In contrast, for samples S1 and S2, the experimentally derived values are significantly higher than the theoretical predictions, despite the fact that their inverse susceptibility curves exhibit fundamentally different behaviors and were fitted over different temperature ranges – a hyperbolic dependence for S1, which does not approach a linear regime, and a nearly linear dependence for S2 (Fig. S1). For the gallium-free sample S4, the theoretical values $C$ and $\mu_{eff}^{theor}$ are instead underestimated, although the qualitative shape of the dependence closely resembles that of S3.

The discrepancies in the extracted parameters may arise because a sufficiently extended linear region in the inverse susceptibility is not reached within the measured temperature range for samples S1 and S4, suggesting that measurements at higher temperatures are required for more reliable Curie-Weiss analysis. For sample S2, a linear dependence is observed at high temperatures, indicating the onset of the paramagnetic regime. However, due to the high measurement temperatures, the experimental data exhibit significant scatter, which may also affect the accuracy of the fit.

Analysis of the Curie–Weiss temperatures $\theta$ reveals that all values are negative, indicating dominant antiferromagnetic interactions in these samples. In the ternary oxides, the magnitude of $\theta$ decreases with increasing gallium content, consistent with the binomial distribution results and suggesting a progressive weakening of magnetic exchange interactions. The most negative $\theta$ value was obtained for sample S2, which showed a clear linear region in the high-temperature regime (Fig. S2). For the other samples, $\theta$ can only be qualitatively assessed due to the lack of well-defined linear behavior.

The anomalously low (in absolute value) Curie–Weiss temperature observed for the Ga-free sample S4 – compared to the other samples and to literature data for related bixbyites [34] – further supports the conclusion that the measured temperature range up to 300 K is insufficient to reach the true high-temperature paramagnetic limit, where $(\chi - \chi_0)^{-1}(T)$ becomes linear. The persistent nonlinearity of $(\chi - \chi_0)^{-1}(T)$ up to $T = 300$ K provides strong evidence for the presence of short-range magnetic correlations well above the observed low-temperature phase transitions (20-40 K). This observation is consistent with earlier findings reported in Ref. [12] for $FeMnO_3$, where magnetic correlations were found to develop at temperatures significantly higher than the spin-freezing transition.

Table 7. Fitting parameters of S1 and S3: Curie-Weiss temperatures $\theta$, experimental and calculated Curie constants $C$ and $C^{theor}$ and the effective magnetic moments $\mu_{eff}^{\exp}$ obtained from the experimental $(\chi - \chi_0)^{-1}(T)$ curves and calculated $\mu_{eff}^{theor}$ from (3).

| Compound | Curie-Weiss law fitting | | | Theoretically predicted values | |
|---|---|---|---|---|---|
| | $\theta$, K | $C$, K·mol·Oe/emu | $\mu_{eff}^{\exp}$, $\mu_B$ | $C^{theor}$, K·mol·Oe/emu | $\mu_{eff}^{theor}$, $\mu_B$ |
| S1 | -473.61 | 7.16 | 7.57 | 6.10 | 6.99 |
| S2 | -564.98 | 6.56 | 7.24 | 5.45 | 6.60 |
| S3 | -212.19 | 4.59 | 6.06 | 4.71 | 6.14 |
| S4 | -99.44 | 4.09 | 5.72 | 6.73 | 7.34 |

Despite extensive investigations aimed at identifying fundamental differences between the paramagnetic samples S1, S3, and S4 (at room temperature) and the ferrimagnetic sample S2, no clear origin for this divergence in magnetic behavior has been established. Furthermore, Mössbauer spectroscopy data – acquired at room temperature – show that all four samples exhibit doublet spectra, indicative of paramagnetic iron. Nevertheless, a clear ferrimagnetic transition is observed in the temperature-dependent magnetization (and consequently in the magnetic susceptibility) of sample S2 at approximately 370 K.

In the course of further investigation, an additional crystal from the same S2 batch was selected for magnetization measurements. The resulting data (at $H = 50$ Oe), converted to magnetic susceptibility, are shown in Fig. 11a alongside the original measurements on the first S2 crystal. The magnetic susceptibility of this second crystal is approximately twice as high at room temperature compared to the initial sample – suggesting the presence of an extrinsic ferro- or ferrimagnetic contribution.

For comparison, Fig. 11a also includes the temperature dependence of the magnetic susceptibility of a single crystal of the spinel $Mn_{1.41}Fe_{1.24}Ga_{0.35}O_4$ [41], which was synthesized in the same experimental series as sample S2 during a study of manganese oxidation state changes as a function of solvent composition [17]. Notably, the ferrimagnetic ordering temperature of this spinel phase is nearly identical to the $T_C$ observed in sample S2. This striking similarity leads us to propose that the ferrimagnetic signal in S2 may be entirely attributed to a minor spinel impurity phase.

This hypothesis is further supported by the fact that such a small impurity would be undetectable by conventional powder X-ray diffraction. Indeed, the XRD pattern of S2 (see

Supplementary Information, "X-ray powder diffraction of S2") is fully consistent with a single-phase bixbyite structure indexed as $Fe_{0.78}Mn_{0.68}Ga_{0.54}O_3$. To illustrate the plausibility of this scenario, the spinel susceptibility signal has been scaled down to 0.5% of its original intensity for comparison with the S2 data. Impurity phases at this level are typically below the detection limit of standard powder XRD techniques.

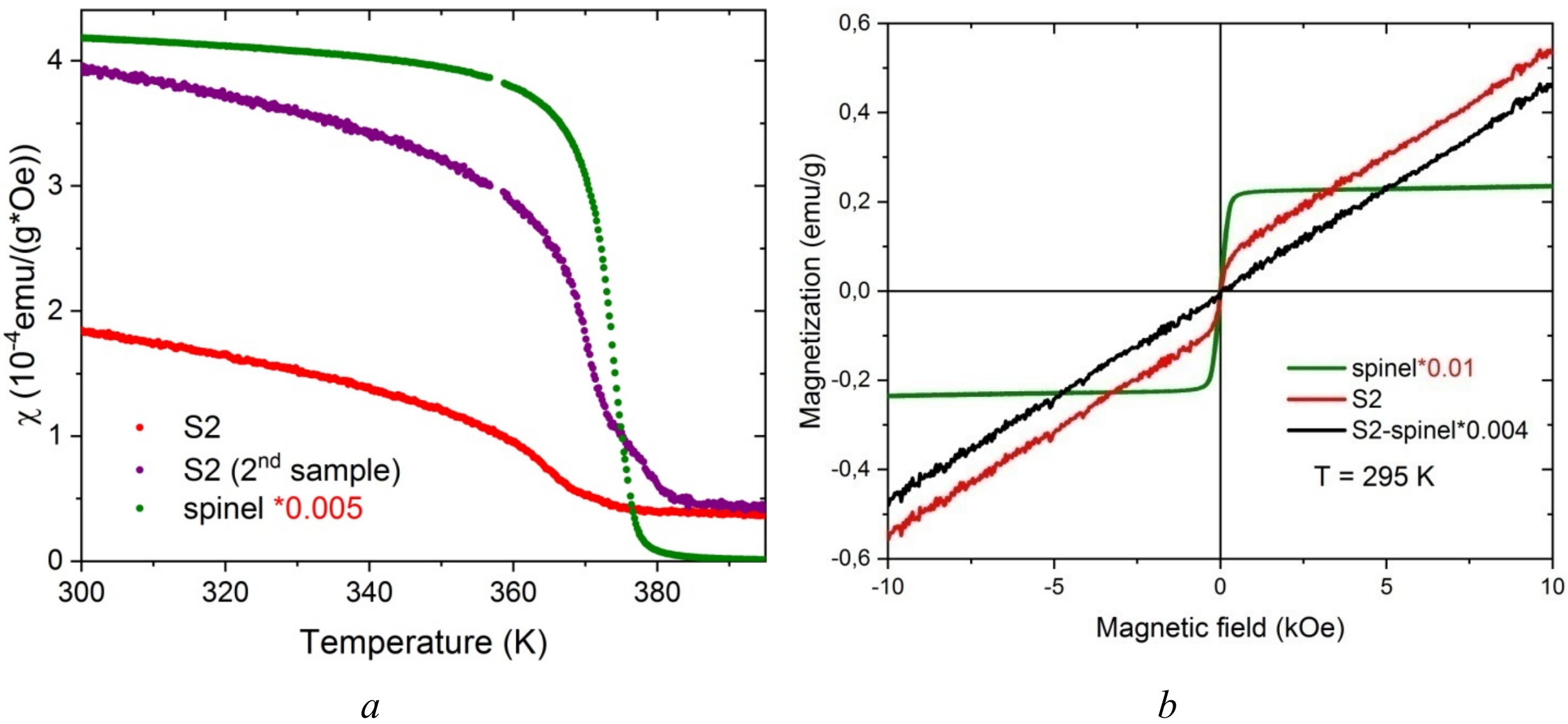

Figure 11. (a) Thermal dependences of the magnetic susceptibility of two S2 samples (red – the same crystal as that used for the magnetization measurements in Section 3.2; purple – another S2 sample) and ferrimagnetic spinel $Mn_{1.41}Fe_{1.24}Ga_{0.35}O_4$ [41] (green) which signal was multiplied by 0.005 for comparison. (b) Field dependences of the magnetization of S2 (red) and spinel $Mn_{1.41}Fe_{1.24}Ga_{0.35}O_4$ (green) whose signal was multiplied by 0.01, measured at $T$ = 295 K. The black curve is the result of the subtraction of spinel curve (multiplied by 0.004) from the S2 $M(H)$ curve.

Fig. 11b compares the field-dependent magnetization of sample S2 (red) and the spinel $Mn_{1.41}Fe_{1.24}Ga_{0.35}O_4$ (green) at $T$ = 295 K, with the spinel signal scaled by a factor of 0.01 for comparison. The curves show excellent agreement near $H$ = 0 Oe, indicating similar low-field magnetic behavior. At fields above 1 kOe, the magnetization of the spinel saturates and remains nearly constant, contributing negligibly to the overall response. In contrast, the magnetization of S2 continues to increase linearly with field, reflecting a dominant paramagnetic contribution from the host bixbyite phase. The close agreement of the low-field M(H) behavior $T$ = 295 K and the coincidence of the anomaly temperatures near $T \approx 370$ K indicate that even a very small amount of a spinel-like phase may account for the observed ferrimagnetic signal. To highlight the impurity contribution to the $M(H)$ signal of S2 the spinel signal (multiplied by 0.004) was subtracted from S2 $M(H)$ curve. The result curve (black in Fig. 11b) is a linear one and reflects only bixbyite S2 contribution.

To support the hypothesis of an impurity-type ferrimagnetic transition in S2, the electron spin resonance (ESR) spectra of this sample and S3 (paramagnetic at room temperature) were measured. This method has high sensitivity for detecting impurity magnetic phases. The ESR spectra were measured on an X-band Bruker Elexsys E580 EPR spectrometer. The ESR spectrum of S2, obtained at $T$ = 300 K, is shown in the insert of Fig. 12a. The spectrum consists of many individual absorption lines, the superposition of which forms a broad envelope. The shape and position of this line are due to the dispersion of the magnetic properties of impurity inclusions in the sample.

As the sample rotates, the positions of individual lines change significantly, but at a temperature of 300 K they always remain in the range of magnetic fields from 2500 Oe to 4200 Oe. Such a significant change in the resonance fields indicates the presence of strong magnetic anisotropy in the impurity, which could be either crystalline magnetic anisotropy or shape anisotropy. It can be assumed that the presence of lines with a *g*-factor less than 2 (resonance field above 3460 Oe) in the spectrum indicates that at least some of the impurity inclusions exhibit pronounced shape anisotropy of the easy-plane type, meaning their lateral dimensions are significantly larger than their thickness. They are most likely oriented parallel to the crystal growth faces.

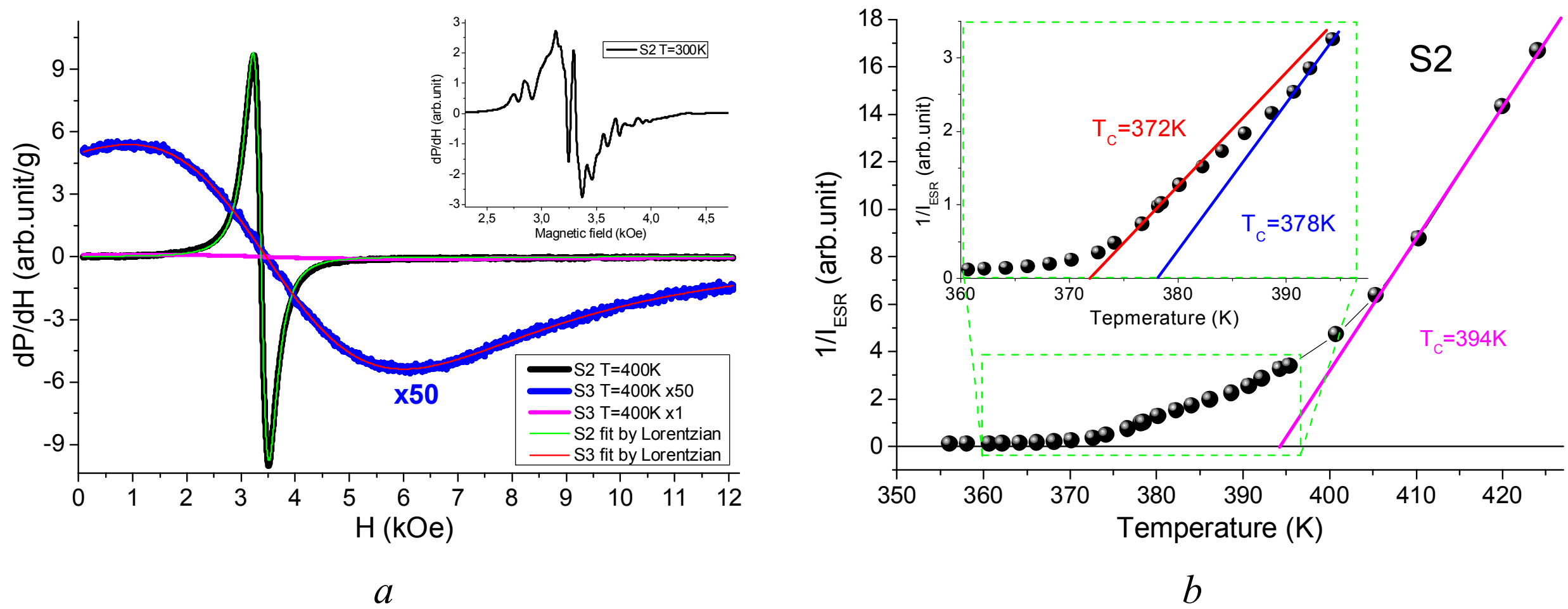

Figure 12. (a) ESR spectra of S2 and S3 at 400 K, the amplitude is normalized to the mass of the sample and ESR spectra of S2 at 300 K (insert); (b) temperature dependences of reciprocal integral ESR intensity of S2 and enlarged section from 360 K to 396 K (insert).

With increasing temperature, individual lines in the ESR spectrum shift toward the center of the broad envelope, while the broad envelope itself moves toward the resonant field of 3460 Oe, which corresponds to a *g*-factor of 2. At a temperature of ~370 K, the individual lines merge into a single line, which is well described by the Lorentzian line. Fig. 12a shows the spectra and the Lorentzian fit for samples S2 and S3 at 400 K. To compare these two samples, their spectra were recorded under identical conditions, and the signal amplitude was normalized to the sample mass. This technique allows for a qualitative assessment of the absorption intensities in different samples, which is more than sufficient in our case. It is clearly seen that the ESR absorption lines from different samples differ significantly in both width and amplitude. The Lorentzian parameters are listed in Table 8.

Table 8. Parameters of ESR spectra of S2 and S3 by Lorentzian: $A$ – amplitude, $\Delta H$ – HWHM, $H_R$ – resonance field , $I_{ESR}$ – integral ESR intensity (2nd integral).

| Sample | $A$ | $\Delta H$, Oe | $H_R$, Oe | $I_{ESR}*10^6$ | *g*-factor |
|---|---|---|---|---|---|
| S2 | 3806 | 252.8 | 3367.7 | 3 | **2.062** |
| S3 | 729 | 4397.2 | 3463.9 | 10 | **1.995** |

By analysis the temperature dependences of reciprocal integral ESR intensity ($1/I_{ESR}$) of S2, which generally follows the Curie-Weiss law in the paramagnetic region the magnetic ordering temperature of the impurity phase can be determined (Fig. 12b). It is evident that the impurity phase is not uniform in terms of magnetic properties. At least three different Curie-Weiss temperatures can be distinguished: 372, 378 and 394 K, which is apparently due to the compositional inhomogeneity of the impurity phase—different Fe/Mn/Ga ratios. This result is consistent with the temperature behavior of the magnetization of sample S2 near the ferrimagnetic phase transition – not one, but two magnetization kinks are observed at temperatures of approximately ~380 and ~275 K (Fig. 11a). Thus, the ESR measurement results confirm the conclusions made earlier based on measurements of magnetization and magnetic susceptibility.

The problem of spinel secondary phase formation is a common for many manganese-contained materials not necessary oxides [42, 43]. From a chemical standpoint, the bixbyite-type oxides $(Fe,Mn,Ga)_2O_3$ and spinel-type $(Fe,Mn,Ga)_3O_4$ differ in the oxidation states of the transition metal cations – particularly manganese. In the spinel structure, each formula unit contains two 3+ cations and one 2+ cation, whereas in bixbyites, all cations are formally 3+ charged. Under uncontrolled synthesis conditions, $Mn^{3+}$ can undergo reduction to $Mn^{2+}$, potentially leading to the formation of a secondary spinel phase enriched in $Mn^{2+}$.

In methods relying solely on the self-flux approach, the primary factor influencing oxidation state changes is reaching the deoxygenation temperature range of the starting $Mn_2O_3$ precursor (900-1100 °C) [17]. However, when using flux-based synthesis with additional non-stoichiometric components – such as solvents or catalysts – chemical interactions with these additives can promote redox processes and oxidation state changes at significantly lower temperatures. Given the inherent instability of the Mn valence state under such conditions, the growth of high-quality, single-phase bixbyite crystals requires not only optimized synthesis protocols but also a comprehensive understanding of crystallization kinetics and redox equilibria.

Initial efforts toward controlled growth of Mn-containing crystals via flux methods [44–47] have revealed strong phase competition, yet also demonstrated strategies to stabilize either $Mn^{2+}$ or $Mn^{3+}$ through careful tuning of oxygen partial pressure, cooling rates, and flux composition.

We propose that the ferrimagnetic behavior previously reported in $(Mn,Fe)_2O_3$ bixbyites at room temperature may similarly arise from trace amounts of a ferrimagnetic spinel impurity, undetectable by conventional X-ray diffraction. Structurally and magnetically, the four bixbyite samples studied here – despite varying compositions – are highly similar, and there is no intrinsic reason to expect a significant difference such as an additional high-temperature ferrimagnetic transition in one of them. The absence of such features in S1, S3, and S4 further supports the extrinsic origin of the anomaly observed in S2.

## 5. Summary and Conclusion

The central focus of this study is the wide variation in reported magnetic properties of $FeMnO_3$-based bixbyites, as observed in different studies conducted under diverse synthesis conditions. To clarify the origin of these discrepancies, single crystals of three $(Fe,Mn,Ga)_2O_3$ samples with varying Mn/Fe/Ga ratios (S1–S3) and one gallium-free sample, $Fe_{0.52}Mn_{1.48}O_3$ (S4), were synthesized using a flux (solution-melt) method and systematically investigated.

A combination of techniques – including powder X-ray diffraction, magnetometry, element-sensitive X-ray absorption spectroscopy (XAS), and Mössbauer spectroscopy – was employed to probe both the structural and magnetic characteristics of the samples. Initially, these methods yielded seemingly contradictory results, mirroring the inconsistencies reported in the literature: one sample (S2) exhibited ferrimagnetic behavior at room temperature in magnetization measurements, while the other three (S1, S3, S4) displayed only low-temperature magnetic transitions at $T$ = 20-40 K, typical of bixbyite-type oxides. The absence of pronounced anomalies in the heat capacity data across this temperature range for all four samples (S1–S4) supports the interpretation that these low-temperature features may arise from spin-glass-like freezing rather than conventional long-range antiferromagnetic ordering.

An attempt to correlate the observed magnetic behavior with cationic positional order in the bixbyite structure – characterized by two inequivalent cation sites (M1: 8*b* and M2: 24*d*) and three types of cations ($Fe^{3+}$, $Mn^{3+}$, $Ga^{3+}$) – did not yield unambiguous conclusions. All samples were found to be positionally disordered, yet consistently exhibited a preferential occupancy of $Mn^{3+}$ cations at the M2 (24*d*) site, regardless of overall composition or cation ratio. This suggests a general trend in site preference rather than a composition-dependent ordering mechanism.

However, a strong dependence of positional order on crystal growth kinetics was revealed. The samples were grown under different cooling rates: $dT/dt$ = 4 °C/day (S1), 8 °C/day (S2), 12 °C/day (S3), and near-isothermal conditions (~0 °C/day) for S4 (though the flux composition also differed significantly). A clear trend emerged: faster cooling rates correlate with increased positional disorder. In the most slowly cooled sample (S1), the Mössbauer spectrum, analyzed using a binomial distribution model, indicates a local iron environment consistent with the two distinct crystallographic sites in the bixbyite structure. Statistically, $Fe^{3+}$ cations in S1 are more likely to be surrounded by other $Fe^{3+}$ ions, and a marked difference in the local environments of the M1 (8*b*) and M2 (24*d*) sites is observed. This suggests that S1 approaches the most thermodynamically stable, cation-ordered state accessible under these conditions. Slower cooling thus promotes cationic ordering, and further reduction in growth rate might lead to even higher degrees of order.

In contrast, samples S2, S3, and S4 exhibit significantly greater disorder. Notably, the room-temperature ferrimagnetic response observed in S2 cannot be attributed to intrinsic ordering, as its cation distribution resembles that of the paramagnetic S3 and S4. Moreover, the Mössbauer spectra of all four samples at room temperature consist of a single doublet, confirming the paramagnetic state of $Fe^{3+}$ ions – consistent with earlier reports [12]. Therefore, the ferrimagnetic behavior of S2 at room temperature cannot be directly explained only by the peculiarities of the positional distribution of cations in the bixbyite structure.

This leads to a key contradiction: while magnetometry on the S2 single crystal suggests ferrimagnetism at room temperature, Mössbauer spectroscopy (performed on a powdered sample) confirms a paramagnetic state. To resolve this discrepancy, additional temperature-dependent magnetization measurements were carried out on a second single crystal of the same composition as S2. The results confirmed the presence of a ferrimagnetic contribution, most likely due to an impurity phase. This phase exhibits a transition temperature $T_C \approx 370$ K – nearly identical to that observed in S2.

We attribute this impurity to a spinel-type phase, likely $(Fe,Mn,Ga)_3O_4$, which can form under non-equilibrium synthesis conditions due to partial reduction of $Mn^{3+}$ to $Mn^{2+}$. This result is confirmed by the measurements and comparison of ESR spectra of S2 ans S3. Even a minute amount (~0.5%) of such a highly magnetically ordered phase can dominate the low-field magnetization signal below $T_C$, especially when the host bixbyite matrix is weakly magnetic. Crucially, such a small impurity fraction is below the detection limit of conventional powder X-ray diffraction, explaining the apparent phase purity of S2.

We therefore propose that many previous reports of room-temperature ferrimagnetism in $FeMnO_3$-based bixbyites may, in fact, be influenced by undetected spinel impurities. The structural and magnetic similarities among the four bixbyite samples studied here – despite differing compositions – further support this conclusion: there is no intrinsic reason for one sample to exhibit a high-temperature magnetic transition while others do not.

In summary, our work highlights the critical importance of phase purity and controlled synthesis in the study of complex magnetic oxides. To obtain physically meaningful results, high-quality, well-characterized single crystals are essential. Achieving such materials requires not only precise control over stoichiometry and growth conditions but also a deep understanding of redox stability, cation diffusion, and phase competition during crystallization.

**Acknowledgements**

The work was carried out within the state assignment of Kirensky Institute of Physics. EDX data, powder X-ray data and the magnetic measurements were obtained on the analytical equipment of the Krasnoyarsk Regional Center of Research Equipment of the Federal Research Center "Krasnoyarsk Science Center SB RAS". The high-temperature magnetic measurements have been supported by the Kazan Federal University Strategic Academic Leadership Program (PRIORITY-2030). This work (M.S.P.) was partially supported by the Ministry of Science and Higher Education of the Russian Federation within the governmental order for SRF SKIF Boreskov Institute of Catalysis (project FWUR-2024-0040).

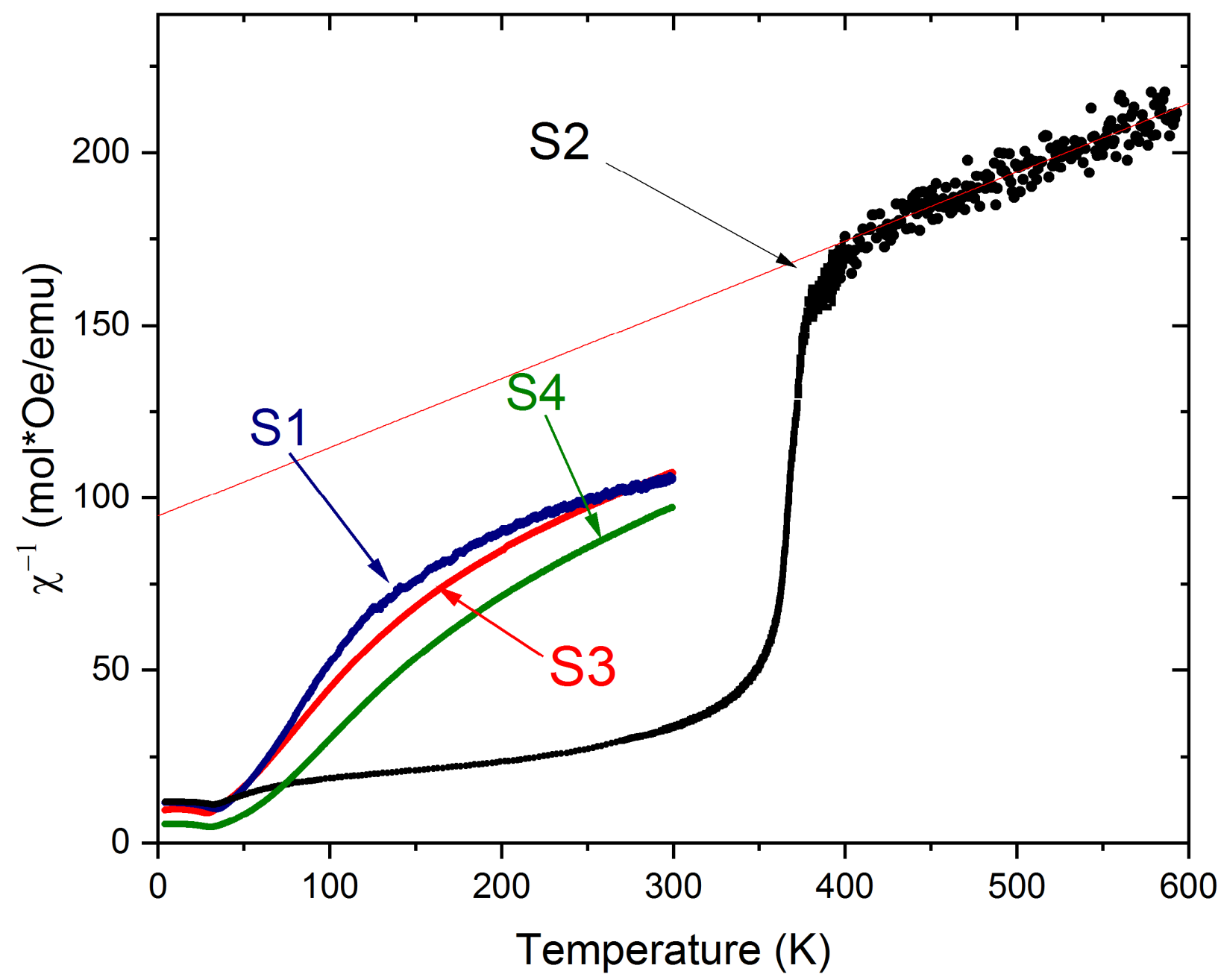

Figure S1. Thermal dependences of inverse magnetic susceptibility of S1 (blue), S2 (black), S3 (red) and S4 (green) samples.

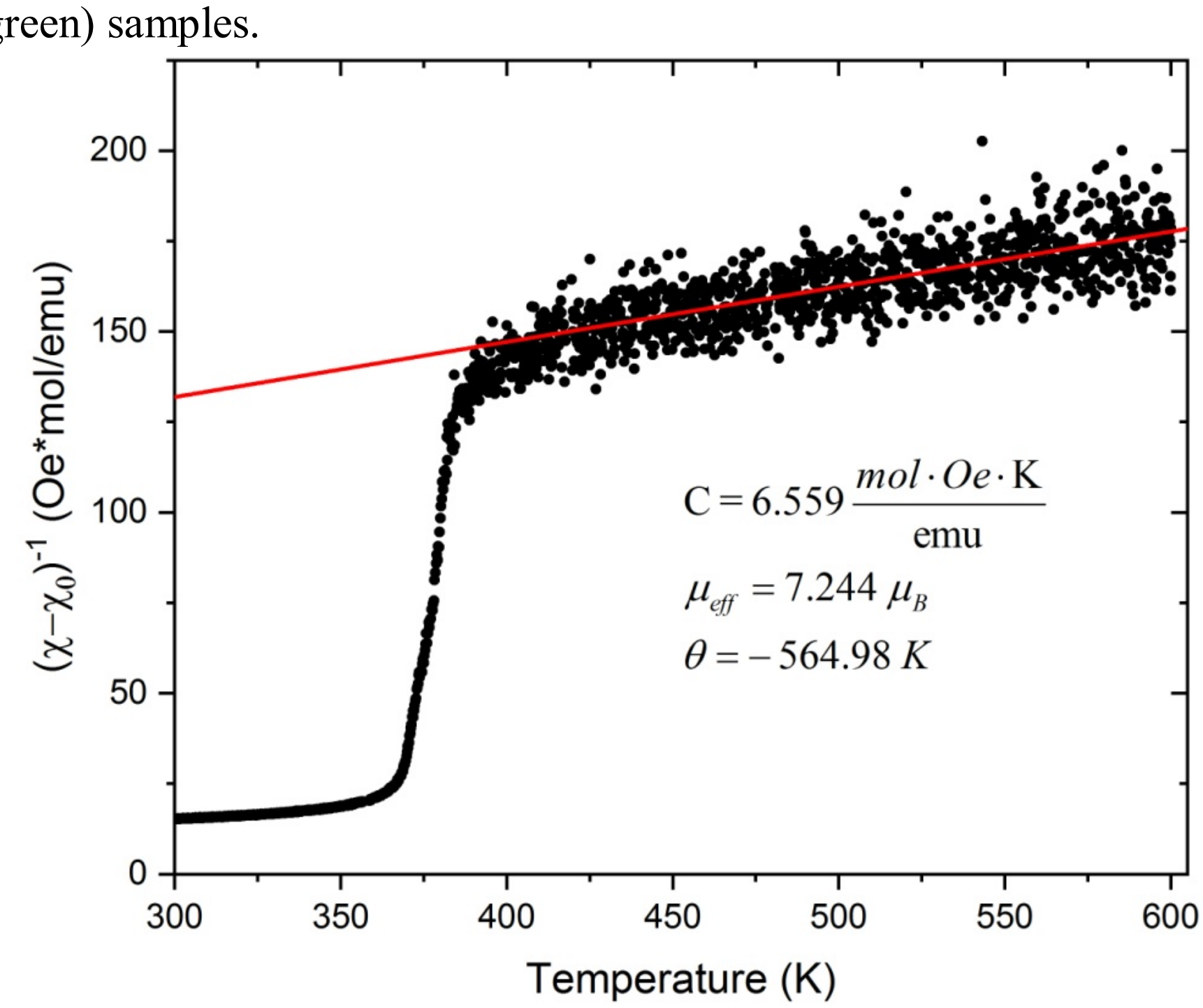

Figure S2. Thermal dependence of inverse magnetic susceptibility of S2 (black) and its fitting by Curie-Weiss law (red line).